\documentclass[aps,prx, superscriptaddress,twocolumn]{revtex4-1}
\usepackage{graphicx}
\usepackage{physics}
\usepackage{float}
\usepackage{amsfonts}
\usepackage{hyperref,xcolor,comment}
\usepackage[normalem]{ulem}

\begin{document}

%\preprint{}

\title{Prestressed elasticity of amorphous solids}


%\author{}
%\thanks{}
%\affiliation{}

\author{Shang Zhang}
 \affiliation{Department of Physics, University of Michigan, Ann Arbor, MI 48109-1040, USA}

\author{Ethan Stanifer}
 \affiliation{Department of Physics, University of Michigan, Ann Arbor, MI 48109-1040, USA}

\author{Vishwas Vasisht}
 \affiliation{Department of Physics, Indian Institute of Technology, Palakkad, India}

\author{Leyou Zhang}
\thanks{Present address: Google Cloud AI, Pittsburgh, PA, 15206, USA} % If there are better ways to show present address, please change this line.
 \affiliation{Department of Physics, University of Michigan, Ann Arbor, MI 48109-1040, USA}

\author{Emanuela Del Gado}
 \affiliation{Department of Physics, Georgetown University, Washington DC, USA}

\author{Xiaoming Mao}
\email{maox@umich.edu}
 \affiliation{Department of Physics, University of Michigan, Ann Arbor, MI 48109-1040, USA}

\date{\today}

\begin{abstract}
Prestress in amorphous solids bears the memory of their formation, and plays a profound role in their mechanical properties, from stiffening or softening elastic moduli to shifting frequencies of vibrational modes, as well as directing yielding and solidification in the nonlinear regime. Here we develop a set of mathematical tools to investigate elasticity of prestressed discrete networks, which disentangles the effects from disorder in configuration and disorder in prestress.   Applying these methods to prestressed triangular lattices and a computational model of amorphous solids, we demonstrate the importance of prestress on elasticity, and 
reveal a number of intriguing  effects caused by prestress, including 
strong spatial heterogeneity in stress-response unique to prestressed solids, 
power-law distribution of minimal dipole stiffness, %long-range nature of transverse stress response, 
and a new criterion to classify floppy modes. 
\end{abstract}

%\keywords{}

\maketitle


%\tableofcontents

\section{Introduction}
Almost all solid materials are stressed. Amorphous solids exhibit quenched residual stress from their preparation process, crystalline solids are stressed by defects and grain-boundaries, and living matter experiences active stress from biological processes. The ubiquity of stressed solids is encapsulated in the fact that the stress tensor $\sigma$, as a $d\times d$ symmetrical matrix in $d$ dimensions, has $d(d+1)/2$ independent degrees of freedom, but the force balance equation, 
\begin{equation}\label{EQ:FB}
    \partial_j \sigma_{ij} =0,
\end{equation}
only poses $d$ constraints, leaving $d(d-1)/2$ unconstrained components in the stress field. In absence of external load, all components have to vanish if the material, and the stress field, have to be homogeneous,  but they can be nonzero at
lengthscales over which heterogeneities or excess constraints are present (``geometric frustration''), 
giving rise to {\it prestress}, (also known as ``residual stress'', ``initial stress''  or ``eigenstress'' in different contexts) \cite{alexander1998amorphous,solids2003}. 

\begin{figure}%[H]
    \centering
    \includegraphics[width=0.5\textwidth]{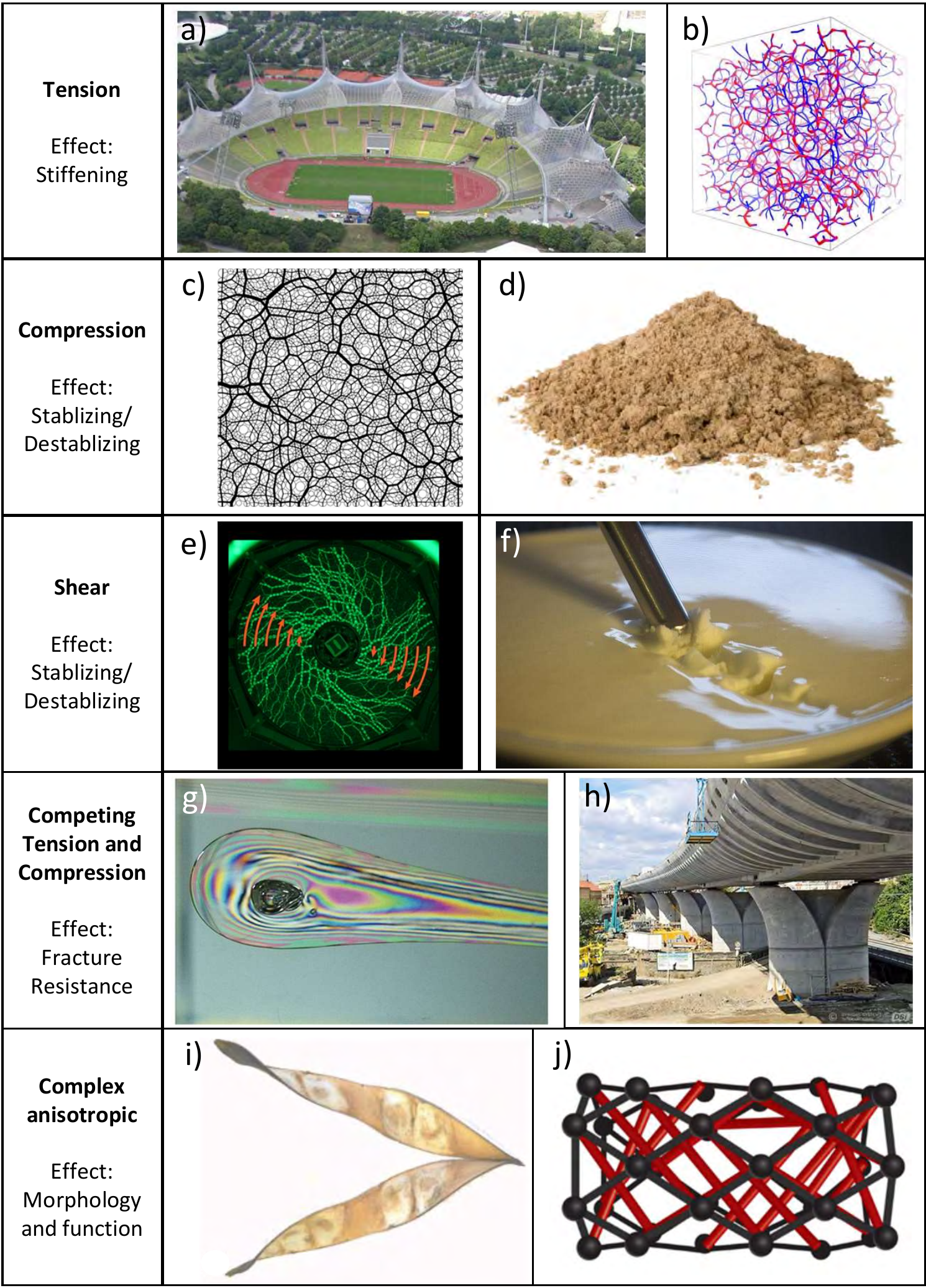}
    \caption{A series of examples of stressed materials including structures under tension [a tensile net architecture in (a) and a model colloidal gel \cite{Bouzid2017elastically}) in (b)], structures under compression [contact network of a jammed packing of repulsive disks~\cite{Bouchaud2002} (c)  and pile of sand (d)], structures under shear [shear jamming of photoelastic discs (e) and an acoustically tunable shear thickening fluid (image taken by Neil Lin) (f)], structures under competing tensile and compressive forces [a Prince Rupert’s drop~\cite{Aben2016} in (g) and a bridge made from prestressed concrete in (h)], as well as systems with more complex modes of stress [a pea pod which uses stress to open~\cite{Armon1726} in (i) and a tensegrity metamaterial \cite{Wen2020} in (j)]. 
    }
    \label{FIG:Intro}
\end{figure}

In structural engineering, prestress is proactively used to modify both stability and load bearing capability of structures, from prestressed reinforced concrete to tensegrity architectures 
[Fig.~\ref{FIG:Intro} (a), (h) and (j)]. In materials, instead, prestresses can emerge spontaneously, as the direct consequence of out-of-equilibrium processes through which they solidify, or of the external load applied during processing.
Prominent examples include isotropic compressive prestress in jammed packings~\cite{Wyart2005} or gels ~\cite{Bouzid2017elastically}, shear prestress in shear jammed granular matter~\cite{Bi2011} and shear thickened dense suspensions~\cite{Sehgal2019}, isotropic prestresses in glasses~\cite{Ballauff2013}, and rich varieties of anisotropic prestress fields in prestressed/tensegrity metamaterials~\cite{Gei2009,Chen2012,Fraternali2014,Meeussen_2020,Merrigan2021} and biological systems~\cite{Armon1726,Wyczalkowski2012,Feng:2018PNAS} [Fig.\ref{FIG:Intro} (b)-(g) and (i)].  
Remarkably, very much in the same way as for buildings and large scale structures, microscopic residual stresses in amorphous solids may strongly affect their strength---stiffen or soften them, and direct how they fail \cite{alexander1998amorphous,zhang2017fiber,Ozawa2018random,vasisht2020emergence,Barlow2020ductile,Berthier2019}. In addition, prestress stores elastic  energy in materials, which then feature a different energy landscape when compared to stress-free materials \cite{debenedetti2001supercooled,PhysRevLett.107.108302,PhysRevX.7.021039,vasisht2021residual,Bhaumike2021role,Richard2020Predicting}.

A deep understanding of the consequences of prestress is therefore key to predict elasticity and material properties in general, but microscopic prestresses can be very elusive, since it is difficult to directly access them in experiments and in most cases only their indirect consequences can be detected. 
It has been shown that, in jamming of frictionless particles, compressive prestress both stabilizes the packing by maintaining particle contacts, and destabilizes it by shifting frequencies of vibrational modes~\cite{Wyart2005,DeGiuli2014}, as well as control the response to external forces~\cite{Ellenbroek2009}. In biological systems such as biopolymer gels and epithelial cell sheets, tensile stress has been shown to provide stability~\cite{Licup9573,feng2015alignment,feng2016nonlinear,noll2017active,hardin2018long,Yan2019,merkel2019minimal,liu2021topological}, and active stress generated by molecular motors has been shown to give rise to sophisticated effects on rigidity~\cite{alvarado2013molecular} and mechanical signal transmission~\cite{ronceray2016fiber}.

However, a general theoretical framework to 
disentangle effects of microstructure and prestress, and map how external load is carried by prestressed amorphous solids, is still missing. Many studies of elasticity of amorphous solids have mainly focused on determining how the disorder in their microstructure affects elasticity \cite{tanguy2002continuum,OHern2003jamming,zaccone2011approximate,liu2010jamming}, whereas the role of prestress is far less understood, so that there is a gap in the theoretical approaches that can rationalize the mechanics of amorphous solids when it comes to prestress.
In a granular solid, the same configuration of particles and the same value of the stress at the boundary are compatible with multiple stress distributions, which correspond to the so-called ``force network ensemble''~\cite{Snoeijer2004,Tighe2008,tighe2010force,Bi2015} and is related to the broader concept of the ``Edwards ensemble''~\cite{edwards1989theory} where both microscopic configuration and stress are allowed to vary for a given boundary stress.
Statistical relations of 
prestress distributions %regarding 
in this type of ensembles have been recently discussed, and force-balance constraints [Eq.~\eqref{EQ:FB}] have been shown to produce %led to 
intriguing long-range correlations~\cite{Mao_2007,Henkes2009,Mao2009,PhysRevE.96.052101,Maier2017emergence, DeGiuli2018,Nampoothiri2020}. Nevertheless, unraveling the role of disordered configurations and disordered prestress in the mechanical properties of amorphous solids has remained difficult both conceptually and computationally.

When considering the stress distributions in a mechanically stable structure, the concept of states of self stress (SSSs) \cite{calladine1978buckminster, pellegrino1986matrix} captures their eigenstates %of stress distributions 
that leave all components of the solid structure in force balance. In discrete mechanical networks, SSSs originate from redundant constraints. SSSs were first introduced in mechanical engineering to describe load bearing abilities of structures, and more recently used by the condensed matter physics community for their %special 
role in characterizing mechanical topological edge states~\cite{sun2012surface,kane2014topological,lubensky2015phonons,mao2018maxwell}. Because SSSs span the linear space of all possible ways a system can carry load, mapping out, and further programming, this SSSs linear space offers a convenient handle to control the mechanical response of materials. Interesting examples of this type include recent work on programming SSSs using topological mechanics to direct buckling and fracturing of metamaterials~\cite{paulose2015selective,zhang2018fracturing}. Intriguingly, SSSs in jammed packings have been shown to exhibit rich spatial structures and a divergent length scale near jamming~\cite{sussman2016spatial}. Some of these findings suggest that SSSs may be the natural tool to rigorously quantify the effect of prestress distributions on material properties, but a direct connection remains to be built.

Here we present a systematic method to investigate how prestress affects the mechanical response of amorphous solids, based on the concept of SSSs, where the configuration and the prestress are allowed to change \emph{independently}, while force-balance is always maintained. As we discuss below, prestress is a linear combination of the SSSs of the stress-free version of the mechanical network, and %prestress 
leads to a new set of SSSs,  capturing the unique, prestress-controlled, mechanical response of the system. By applying this method to a prestressed triangular lattice model and a computational model of amorphous solids, we show that prestress greatly enlarges the linear space of stress-bearing states, and that a number of intriguing effects arise where prestress strongly controls the mechanical response, both at the local and the global level. 

The general method we introduce here is applicable to a wide range of systems,  ordered and disordered, where prestress affects elasticity. It provides an efficient computational algorithm for finding the stress distribution when the system is under any load, as well as offering a platform to develop a field-theoretic treatment of prestressed elasticity of amorphous solids. Because this method allows the microstructure and the prestress field to vary separately, it offers a pathway to investigate how amorphous solids {evolve as well as develop memory under stress without changing their configuration}. We demonstrate the method using a triangular lattice model with varying prestress, and test it in amorphous configurations of compressed repulsive particles, obtained through numerical simulations. In such model solids, we show how prestress determines their response to both macroscopic shear strain and local dipole forces, and show that they display qualitatively different behaviors from unstressed spring networks with the same geometry. Finally, for the model solids obtained through computer simulations, we use the new method to study the dependence of the stress-bearing ability on the preparation protocol, which changes the microscopic prestress distribution, as well as various signatures of the spatial evolution of the stress under a shear strain.

Overall, the new method proposed here is an ideal candidate to investigate the nature of rigidity transitions in prestressed systems, and furthermore, 
yielding and shear thickening/thinning systems such as granular matter or jammed packings and dense suspensions, to potentially shed new light on the dynamical interplay between stress and geometry in these complex materials.

The paper is organized as follows.  In Sec.~\ref{SEC:Prestress} we review the fundamental theories of prestressed mechanical networks.  In Sec.~\ref{SEC:SSS} we introduce a new formulation of equilibrium and compatibility matrices in prestressed networks, and discuss how they control both zero modes and stress response.  In Sec.~\ref{SEC:Triangular} we use a prestressed triangular lattice model to illustrate how prestress controls the vibrational modes and the stress response.  In Sec.~\ref{SEC:Amorphous} we apply these methods to analyze the stress response of a computational model of prestressed amorphous solids, and Sec.~\ref{SEC:Conclusion} contains a summary and outlook.

%%%%%%%%%%%%%%%%%%%%%%%%%%%%
%%%%%%%%%%%%%%%%%%%%%%%%%%%%
%%%%%%%%%%%%%%%%%%%%%%%%%%%%
\section{Prestressed mechanics}\label{SEC:Prestress}
%%%%%%%%%%%%%%%%%%%%%%%%%%%%
%%%%%%%%%%%%%%%%%%%%%%%%%%%%
%%%%%%%%%%%%%%%%%%%%%%%%%%%%
In this section we briefly review mechanics of prestressed  networks, their continuum elasticity limit, as well as how prestress affects rigidity of discrete networks in general.
\subsection{Prestressed mechanical networks}
We consider a discrete network of point-like particles connected by pairwise central-force potentials (bonds).  When this network is deformed, particle $\ell$, which was originally  at $\vec{R}_{\ell,0}$ in the reference state, undergoes a displacement $\vec{u}_{\ell}$ to a new position
\begin{equation}\label{EQ:Ru}
    \vec{R}_{\ell} = \vec{R}_{\ell,0}+\vec{u}_{\ell}.
\end{equation}
The change of the elastic energy $V_b$ associated with this deformation, of a bond $b$ connecting particles $\ell,\ell'$, is $\delta V_b =V_b(\vert \vec{R}_{b} \vert) - V_b(\vert \vec{R}_{b,0} \vert)$, 
where $\vec{R}_{b,0}\equiv \vec{R}_{\ell,0}-\vec{R}_{\ell',0}$, $\vec{R}_{b}\equiv \vec{R}_{\ell}-\vec{R}_{\ell'}$ are the vectors along the bond in the reference state and the deformed state respectively.

This change in energy
can be expanded to quadratic order in $u$ [using Eq.~\eqref{EQ:Ru}]
%in \eqref{EQ:deltaV}] as 
\begin{equation}\label{EQ:dVb}
    \delta V_b 
    =\frac{ V_b''}{2} 
     |e_b^{\parallel}|^2 
    + \frac{V_b'}{2\vert \vec{R}_{b,0} \vert } |e_b^{\perp}|^2 ,
\end{equation}
where  the derivatives are taken at $|\vec{R}_{b,0}|$, and $\vec{e}_{b} \equiv \vec{u}_{\ell}-\vec{u}_{\ell'}$ 
is the difference of displacement vectors of the two particles connected by  bond $b$, and
\begin{equation}\label{EQ:e2}
    e_b^{\parallel} \equiv 
    \hat{R}_{b,0}\hat{R}_{b,0}\cdot \vec{e}_{b} ,\quad
    e_b^{\perp} \equiv (\mathbb{I}-\hat{R}_{b,0}\hat{R}_{b,0}) \cdot \vec{e}_{b}
\end{equation}
are its components parallel and perpendicular to the original bond direction $\hat{R}_{b,0}\equiv \vec{R}_{b,0}/|\vec{R}_{b,0}|$, respectively.

The derivatives of this potential,
\begin{align}
    V_b'' (|\vec{R}_{b}|) \equiv 
    \frac{d^2 V_b(|\vec{R}_{b}|)}{d|\vec{R}_{b}|^2} =k_b ,
    \nonumber \\
    V_b' (|\vec{R}_{b}|) \equiv 
    \frac{d V_b(|\vec{R}_{b}|)}{d|\vec{R}_{b}|}
    = t_{b,p},
\end{align}
correspond to the spring constant $k_b$ and the pretension $t_{b,p}$ (tension of the bond in the reference state, where $p$ stands for ``prestress'') of bond $b$.

In a stress-free network, all bonds are at their rest length $R_{b,R}$ and thus $V_b'(|\vec{R}_{b,0}|)=0$, leaving only the $V_b''$ term in Eq.~\eqref{EQ:dVb}.  Vibrational modes of disordered networks of this type have been extensively studied, yielding a rich set of interesting phenomena including quasilocalized modes, anomalies of density of states at low frequencies, etc.~ \cite{tanguy2002continuum,OHern2003jamming,Wyart2005,wyart2008elasticity,liu2010jamming,chen2011measurement,lerner2012unified,Mosayeby2014soft,stanifer2018simple}. 

When a network is prestressed, $V_b'(|\vec{R}_{b,0}|)\ne 0$, and both terms contribute to the elastic energy, increasing the number of  microscopic constraints, as we discuss in detail in Sec.~\ref{SEC:SSS}.  Interestingly, in general the sign of $t_{b,p}$ can be either positive (tension) or negative (compression).  
In the case of $t_{b,p}>0$, the prestress term contribute with another complete square term to the elastic energy, clearly stabilizing the system. In the case of $t_{b,p}<0$, naively, the prestress term appears to be unstable.  However, because $e_b^{\perp}$ are not variables independent of $e_b^{\parallel}$ (of other bonds), the stability of the network needs to be analyzed in terms of the collective modes, which we discuss more in Sec.~\ref{SEC:Rigidity}.

Here we assumed that the network in the reference state is in \emph{force balance}, i.e., the total force on each particle vanishes. As a result, there is no $\mathcal{O}(e)$ term in the expansion of the elastic energy. Note that \emph{the force-balance condition and the stress-free condition are two distinct conditions}, where the later means $V_b'(|\vec{R}_{b,0}|)=0$ on all bonds, and is a much more stringent requirement than the force-balance condition.  We will revisit this distinction in continuum elasticity in Sec.~\ref{SEC:Continuum}.

\subsection{Continuum elasticity with prestress}\label{SEC:Continuum}
The discrete theory discussed above can be rigorously linked to continuum elasticity using the relation between the discrete nonlinear strain $v_b$ of bond $b$ defined as~\cite{born1955dynamical,didonna2005nonaffine} 
\begin{equation}
    v_{b} \equiv
    \frac{1}{2}\left(
    |\vec{R}_{b}|^2 - |\vec{R}_{b,0}|^2
    \right)
\end{equation}
and the (continuum) nonlinear right Cauchy-Green strain tensor (repeated indices are summed over)
\begin{equation}
    \epsilon_{ij}(\vec{x}) \equiv
    \frac{1}{2}\left(
    \partial_i u_j + \partial_j u_i +\partial_i u_l\partial_i u_l
    \right)
\end{equation}
where the relation reads 
\begin{equation}\label{EQ:vbCont}
   v_b = \vec{R}_{b,0} \cdot \epsilon_{ij}(\vec{x}) \cdot \vec{R}_{b,0} .
\end{equation}
This follows from the definition of the nonlinear strain tensor, and  we have taken the continuum limit by assuming that the spatial variation of the strain field is small over the scale of the particles and bonds, so that the deformation of bond $b$ is determined by the strain at its location $\vec{x}$. 

By expanding  the bond length $|\vec{R}_{b}|$ in terms of $v_b$,
% \begin{equation}
%     |\vec{R}_{b}| = |\vec{R}_{b,0}|
%     \left\lbrack
%     1+ \frac{v_b}{|\vec{R}_{b,0}|^2}
%     - \frac{v_b^2}{2|\vec{R}_{b,0}|^4}
%     + \mathcal{O} \left(\frac{v_b}{|\vec{R}_{b,0}|^2} \right)^3
%     \right\rbrack
% \end{equation}
we can rewrite the expansion of the elastic energy of bond $b$ as
\begin{equation}\label{EQ:dVbvb}
    \delta V_b 
    =
    V_b' \frac{v_b}{|\vec{R}_{b,0}|} +\frac{1}{2}\left( V_b''-\frac{V_b'}{|\vec{R}_{b,0}|}\right) 
    \left( \frac{v_b}{|\vec{R}_{b,0}|} \right)^2 .
\end{equation}
This is equivalent to the expansion in Eq.~\eqref{EQ:dVb}, by recognizing that 
\begin{equation}\label{EQ:vbeb}
    v_b 
    = \vec{R}_{b,0} \cdot \vec{e}_b +\frac{1}{2}\vec{e}_b \cdot \vec{e}_b
    = |\vec{R}_{b,0}| e_{b}^{\parallel} + \frac{(e_{b}^{\parallel})^2  +(e_{b}^{\perp})^2 }{2}.
\end{equation}
The combination of Eqs.~\eqref{EQ:vbCont} and \eqref{EQ:dVbvb} allows us to  write the bond energy change in terms of the strain tensor.  

The elastic energy of the whole system can be taken to the continuum limit by converting the sum over all bonds to an integral over space
\begin{equation}\label{EQ:cell}
 E= \frac{1}{2} \sum_{\ell} \sum_{\ell'}  V_{b=\langle \ell,\ell'\rangle} =\frac{1}{2} \int d^d \vec{x} \frac{1}{v(\vec{x})} \sum_{\ell'} V_{b=\langle \ell,\ell'\rangle}
\end{equation}
where the space is Voronoi tessellated according to the particles, and $v(\vec{x})$ is the volume of the Voronoi cell at $\vec{x}$, and the overall factor of $1/2$ comes from the fact that every bond is counted twice. The sum $\sum_{\ell'}$ in the continuum limit represent the sum over the bonded neighbors of the particle at $\vec{x}$.  Using this formulation, the elastic energy  of the system can be written in the conventional form
\begin{equation}\label{EQ:ECont}
  E = \int d^d \vec{x} \left\lbrack
  \frac{1}{2} K_{ijkl} (\vec{x}) \epsilon_{ij} (\vec{x})
  \epsilon_{ij} (\vec{x})
  + \sigma_{p,ij} (\vec{x}) \epsilon_{ij} (\vec{x})
  \right\rbrack
\end{equation}
where the local elastic-modulus tensor $K_{ijkl} (\vec{x})$ and prestress field $\sigma_{p,ij} (\vec{x})$ are determined from the discrete network by
\begin{align}\label{EQ:Ksigma}
  K_{ijkl} (\vec{x}) =&  \frac{1}{2 v(\vec{x})} \sum_{\ell'} \left( V_b''|\vec{R}_{b,0}|^2 -V_b'|\vec{R}_{b,0}|\right)
  \nonumber\\
  & \quad \cdot \hat{R}_{b,0,i}\hat{R}_{b,0,j}\hat{R}_{b,0,k}\hat{R}_{b,0,l}, \nonumber\\
  \sigma_{p,ij} (\vec{x}) =&  \frac{1}{2 v(\vec{x})} \sum_{\ell'} (V_b'|\vec{R}_{b,0}|) \hat{R}_{b,0,i}\hat{R}_{b,0,j},
\end{align}
where the sum $\ell'$ is over all connected neighbors of particle $\ell$, which is at position $\vec{x}$, and $b=\langle \ell,\ell'\rangle$.   
From this relation, it is clear that the prestress field $\sigma_p(\vec{x})$ in the continuum theory comes from pretensions on bonds which are not at their rest length in the discrete network, whereas the elastic-modulus tensor depends on both the spring constants and the tensions.

The body-force in this continuum theory
\begin{align}\label{EQ:force}
  f_i(\vec{x})= \partial_j  \sigma_{p,ij}(\vec{x})
  = \frac{1}{2 v(\vec{x})} \sum_{\ell'} (V_b'|\vec{R}_{b,0}|) \hat{R}_{b,0,i}
\end{align}
corresponds to the total force on each particle (normalized by the volume of the Voronoi cell) in the discrete network.  Thus, the force-balance conditions in the discrete network and the continuum theory are indeed the same condition.

It is often useful to write this continuum theory in a quadratic expansion in terms  of the displacement field $\vec{u}(\vec{x})$,
\begin{align}\label{EQ:EContu}
  E = &\int d^d \vec{x} \Big\lbrack
  \frac{1}{2} K_{ijkl} (\vec{x}) \partial_i u_j (\vec{x})
  \partial_j u_i (\vec{x}) \nonumber\\
  &
  + \sigma_{p,ij} (\vec{x}) \partial_i u_l (\vec{x})\partial_j u_l (\vec{x})
  \Big\rbrack
\end{align}
where we have used the symmetry of $K_{ijkl}$ and the force balance condition which eliminates $\mathcal{O}(u)$ terms in this elastic energy. 

\subsection{Prestressed rigidity}\label{SEC:Rigidity}
The concept of ``rigidity'' has been a central theme in the discussion of mechanics of soft materials. In general, rigidity can be interpreted from two viewpoints, (i) a microscopic one, where rigidity is attributed to the vanishing of floppy modes (i.e. modes of deformation that cost no elastic energy)~\cite{maxwell1864calculation,tay1984recent}, and (ii) a macroscopic one, where rigidity is defined as the emergence of a spanning rigid cluster that can transmit stress~\cite{jacobs1995generic}.  Remarkably, these two rigidity criteria coincide for the jamming transition of frictionless repulsive particles~\cite{liu2010jamming}.

In this section we focus on the first viewpoint and discuss how prestress changes floppy modes. In Sec.~\ref{SEC:SSS} we discuss how prestress affects the ability of the system to carry additional loads (second point of view) with the help of the concept of SSSs.

Rigidity of stress-free discrete mechanical networks is often analyzed via the comparison between the numbers of  constraints and degrees of freedom~\cite{maxwell1864calculation,jacobs1995generic}. This is well summarized by the Maxwell-Callandine index theorem~\cite{calladine1978buckminster,sun2012surface}.
Rigorously speaking, when this theorem determines that the number of floppy modes of a system vanishes, it implies that the system is ``first-order rigid'', where the energy increases quadratically for all deformations.  
By contrast, in a ``second-order rigid'' system the leading order expansion of the energy with respect to some deformations is of higher order such as cubic or quartic~\cite{connelly1992stability,connelly1996second}.

\begin{figure*}%[H]
    \centering
    \includegraphics[width=0.8\textwidth]{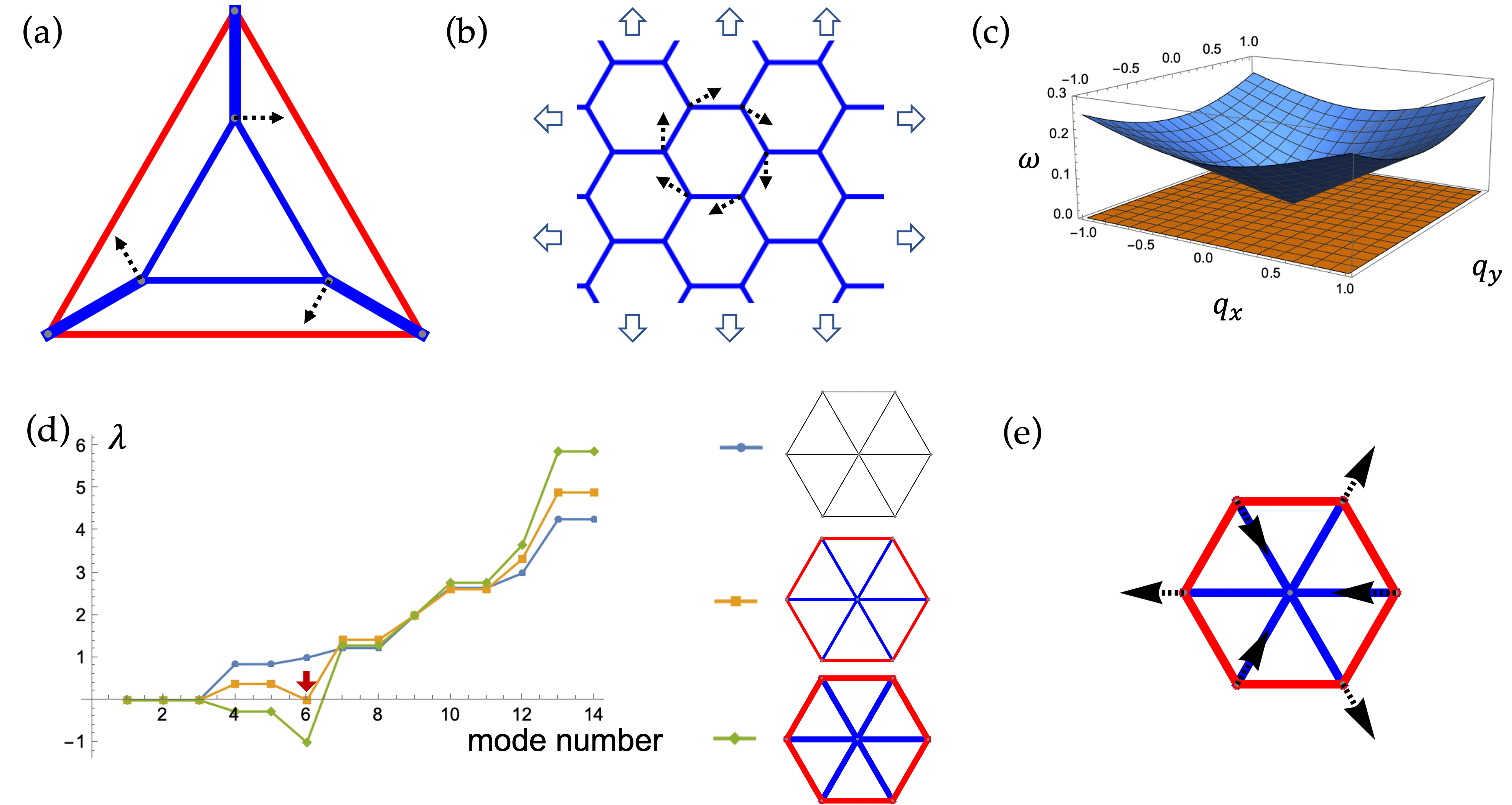}
    \caption{ Prestressed rigidity. (a) A mechanical network of concentric similar triangles~\cite{connelly1992stability}.  When stress-free, this network has one floppy mode (black dashed arrows).  When prestressed, this floppy mode is eliminated.    Blue (red) bonds are under tension (compression).  The thickness of the bond is proportional to the magnitude of the prestress.  (b) A honeycomb lattice exhibits an extensive number of floppy  modes when stress-free (one example shown as dashed arrows).  When prestressed with equal tension on all bonds (blue), these floppy modes are eliminated. (c) Lowest phonon band of the honeycomb lattice when stress-free (orange) and when stretched with tensile prestress as shown in (b) (blue). 
    (d) Eigenvalues $\lambda$ of the dynamical matrix of a mechanical network (right).  When prestress on the network increase beyond a threshold, one normal mode (e) becomes unstable (mode number $6$ as shown by the red arrow), demonstrating an example of type B ZMs discussed in Sec.~\ref{SEC:QCZM} [black dashed arrows in (e)].
    }
    \label{FIG:rigidity}
\end{figure*}

How does prestress change this paradigm of rigidity? It was pointed out in Ref.~\cite{connelly1992stability,connelly1996second} that prestress can rigidify second-order rigid systems and make all their modes first-order rigid.  
Interestingly, this prestress stabilization effect can even take place in systems which are underconstrained  according to the (stress-free) Maxwell counting. One simple example is the honeycomb lattice, which has coordination number $z=3$ and  is thus significantly below the Maxwell point of $z=2d$ ($d$ being the spatial dimension) where degrees of freedom and constraints balance.  When a honeycomb lattice is under tension, all modes are lifted to finite frequency (except for trivial translations). It is worth noting that this example does not violate the theorems in Ref.~\cite{connelly1996second}: although underconstrained, floppy modes of the honeycomb lattice under periodic boundary conditions are in fact second order rigid. In Fig.~\ref{FIG:rigidity} we show examples of how networks become rigid by prestress.

This effect has been studied in various disordered mechanical networks from polymer gels to jamming of particles and biological tissues~\cite{wyart2008elasticity,Licup9573,feng2016nonlinear,vermeulen2017geometry,merkel2019minimal,damavandi2021energetic}. In these models, the underconstrained network typically exhibit an extensive number of floppy modes before the application of stress. When certain types of strain (tensile or shear) is applied to the network, a system spanning SSS emerges and becomes stressed, providing (macroscopic) rigidity and thus elastic moduli. This prestress also rigidifies floppy modes in the network, as the geometry of the network evolves and these modes become second order rigid.

Interestingly, by increasing the magnitude of the prestress, a stable network can also destabilize. This happens when the negative terms in Eq.~\eqref{EQ:dVb}, due to compressed bonds, compete with the positive terms and creates negative eigenvalues in the dynamical matrix. Some examples of this effect are also shown in Fig.~\ref{FIG:rigidity}.
As we introduce our new mathematical tools in Sec.~\ref{SEC:SSS}, we discuss this effect in more detail.

%%%%%%%%%%%%%%%%%%%%%%%%%%%%%
%%%%%%%%%%%%%%%%%%%%%%%%%%%%%
%%%%%%%%%%%%%%%%%%%%%%%%%%%%%
%%%%%%%%%%%%%%%%%%%%%%%%%%%%%

\section{States of self-stress in stressed elasticity}\label{SEC:SSS}

Now we introduce a new $\mathbb{Q},\mathbb{C}$ (equilibrium and compatibility) decomposition for the mechanics of prestressed systems, that can be used to analyze their SSSs and zero modes (ZMs).  We first briefly review this decomposition for stress-free systems, and then discuss the more general case of prestressed systems. We also develop a general formulation of these SSSs to compute mechanical response to an external load, including both macroscopic strain and local forces.

\subsection{States of self-stress in stress-free systems}
The elastic energy of a stress-free network of $N$ point-like particles connected by $N_b$ central-force springs in $d$ dimensions can be written to  quadratic order in $u$ using the dynamical matrix $\mathbb{D}^{\parallel}$ as
\begin{equation}\label{EQ:ESF}
  E = \sum_{b=1}^{N_b} \delta V_b
  = \sum_{b=1}^{N_b} \frac{k_b}{2} (e_b^{\parallel})^2
  = \frac{1}{2} \langle u | \mathbb{D}^{\parallel} | u \rangle 
\end{equation}
where the inner product is taken in the $Nd$ dimensional space for particle displacements $u$, and the superscript ``$\parallel$'' on the dynamical matrix signifies that this dynamical matrix describes a stress-free system where only $e^{\parallel}$ enters the elastic energy.  The more general form for prestressed systems will be discussed in Sec.~\ref{SEC:QCstress}.  
This quadratic form can be decomposed into two steps using the equilibrium and compatibility matrices, defined as
\begin{align}
\mathbb{C}^{\parallel} | u \rangle &= | e^{\parallel} \rangle \label{EQ:CU}
\\
\mathbb{Q}^{\parallel}  | t^{\parallel} \rangle &= - | f \rangle .
\label{EQ:QT}
\end{align}
Here  $| e^{\parallel} \rangle$ and $| t^{\parallel} \rangle$ represent the extension and tension of every bond, which are both $N_b$ dimensional vectors, and the superscript ``$\parallel$'' indicated that they are along the bond direction. $| u \rangle$ and $| f \rangle$ represent the displacement and total force on every site, which are both $Nd$ dimensional vectors.
As a result, the equilibrium matrix $\mathbb{Q}$ has the dimension $Nd \times N_b$, and the compatibility matrix $\mathbb{C}$ has the dimension $N_b \times Nd$. The two matrices are in fact transpose of one another, $\mathbb{Q}=\mathbb{C}^T$.
Note we take the convention that $t^{\parallel}>0$ denotes tension and $t^{\parallel}<0$ denotes compression.  This is consistent with the minus sign of Eq.~\eqref{EQ:QT}, as $f$ is the force \emph{exerted by the bonds in the network}on the sites.  

Using these relations in Eq.~\eqref{EQ:ESF}, it is straightforward to see that 
\begin{equation}
    \mathbb{D}^{\parallel} = \mathbb{Q}^{\parallel} \cdot \mathbb{K}^{\parallel} \cdot \mathbb{C}^{\parallel}
\end{equation}
where $\mathbb{K}^{\parallel}$ is a diagonal matrix that contains all the spring constants $k_b$. In the sign convention we use here,  Hookian's law on the springs takes the form $t^{\parallel}=\mathbb{K}^{\parallel}e^{\parallel}$, as extension causes tension on the springs.

The null space of  $\mathbb{Q}^{\parallel}$ is the set of tensions, called SSSs, that produce no forces at any site,
\begin{align}
0 &= \mathbb{Q}^{\parallel}| t^{\parallel}_{\text{SSS}} \rangle .
\end{align}
The null space of  $\mathbb{C}^{\parallel}$ is the set of site displacements, called ZMs, that produce no changes in bond lengths,
\begin{align}
0 &= \mathbb{C}^{\parallel}
| u_{\text{ZM}} \rangle .
\end{align}
Floppy modes are a subset of ZMs excluding trivial rigid body motions of the whole system, and have been extensively studied in soft matter systems, due to their obvious significance as capturing deformations with no cost of elastic energy. SSSs have only recently been explored in soft matter, but also show great potential in characterizing stress-bearing structures in mechanical networks  \cite{lubensky2015phonons,sussman2016spatial}.

Applying rank-nullity theorem on $\mathbb{Q},\mathbb{C}$ matrices leads to the Maxwell-Calladine index theorem~\cite{calladine1978buckminster,sun2012surface}
\begin{align}
    N_0 - N_S=Nd-N_b,
\end{align}
where $N_0,N_S$ are the numbers of ZMs and SSSs.

These concepts have found wide applications recently in the new field of topological mechanics, as $t^{\parallel}_{\text{SSS}}$ and $u_{\text{ZM}}$ can become topologically protected modes in Maxwell lattices and networks~\cite{kane2014topological,lubensky2015phonons,rocklin2017transformable,zhang2018fracturing,PhysRevLett.120.068003,PhysRevLett.121.094301,mao2018maxwell,PhysRevX.9.021054,PhysRevB.101.104106,PhysRevLett.124.207601}.

\subsection{States of self-stress in prestressed systems}\label{SEC:QCstress}
Prestress on a force-balanced mechanical network can always be viewed as  ``exciting'' an existing SSS in the ``stress-free version'' of the same network (which is generated by turning off stress in the prestressed network but keeping exactly the same geometry---requiring to redefine the rest-length of the springs). Thus, the stress-free version of a prestressed network must exhibit a SSS in the first place.

With prestress, the elastic energy includes both $e_b^{\parallel}$ and $e_b^{\perp}$ terms as analyzed in Eq.~\eqref{EQ:dVb},
\begin{align}\label{EQ:ES}
  E =& \sum_{b=1}^{N_b} \delta V_b
  = \sum_{b=1}^{N_b} 
  \left\lbrack\frac{k_b}{2} (e_b^{\parallel})^2 + \frac{t_{b,p}}{2\vert \vec{R}_{b,0} \vert } |e_b^{\perp}|^2 
  \right\rbrack \nonumber\\
  =& \frac{1}{2} \langle u | (\mathbb{D}^{\parallel}+\mathbb{D}^{\perp}) | u \rangle .
\end{align}
Similar to the stress-free case, using the fact that this elastic energy consists only of complete square terms, this dynamical matrix can also be decomposed into $\mathbb{Q},\mathbb{C}$ matrices,
\begin{equation}\label{EQ:DM}
    \mathbb{D} = \mathbb{Q} \cdot \mathbb{K} \cdot \mathbb{C},
\end{equation}
where
\begin{align}\label{EQ:Ctotal} 
    \mathbb{C} | u \rangle &= 
\begin{pmatrix}
e^{||}\\
e^{\perp}
\end{pmatrix}
\equiv| e \rangle ,
\end{align}
defines the new $\mathbb{C}$ matrix and $\mathbb{Q}=\mathbb{C}^T$.  
Here we have defined  the new $N_bd$ dimensional $e$ vector which contains one component from $\parallel$ and $d-1$ from $\perp$ for each of the $N_b$ bonds.
It is worth noting here that the compatibility matrix $\mathbb{C}$ is now $N_bd \times Nd$ dimensional instead of $N_b \times Nd$ dimensional, because this $\mathbb{C}$ matrix maps the $Nd$ dimensional displacement vector of the network into $e_b^{\parallel}$ and $e_b^{\perp}$ for each bond $b$.  At the same time, the equilibrium matrix $\mathbb{Q}$ is $Nd \times N_bd$ dimensional. The spring constant matrix is a  $N_bd \times N_bd$ diagonal matrix with spring constant $k_b$ for the $\parallel$ terms and pre-stress $t_{b,p}$ for the $\perp$ terms,
\begin{align}
\mathbb{K} = 
\begin{pmatrix}
k_b &  0\\
0 & \frac{t_{b,p} }{ |\vec{R}_{b,0}|}
\end{pmatrix},
\end{align}
where there are $d$ blocks of $N_b$ dimensional diagonal matrices (one longitudinal and $d-1$ transverse).  

These new $\mathbb{Q},\mathbb{C}$ matrices  describe the mapping between the ($Nd$ dimensional) degrees of freedom space and the (now $N_b d$ dimensional) constraint space.  In parallel with Eq.~\eqref{EQ:Ctotal}  we have
\begin{align}
\mathbb{Q} 
\begin{pmatrix}
t^{||}\\
t^{\perp}
\end{pmatrix}
&\equiv \mathbb{Q} 
| t \rangle
= - | f \rangle
\label{EQ:Qtot}
\end{align}
Correspondingly,  $t^{\parallel}$ and $t^{\perp}$ are the parallel and perpendicular components of $t$.  

\begin{figure}[H]
    \centering
    \includegraphics[width=0.5\textwidth]{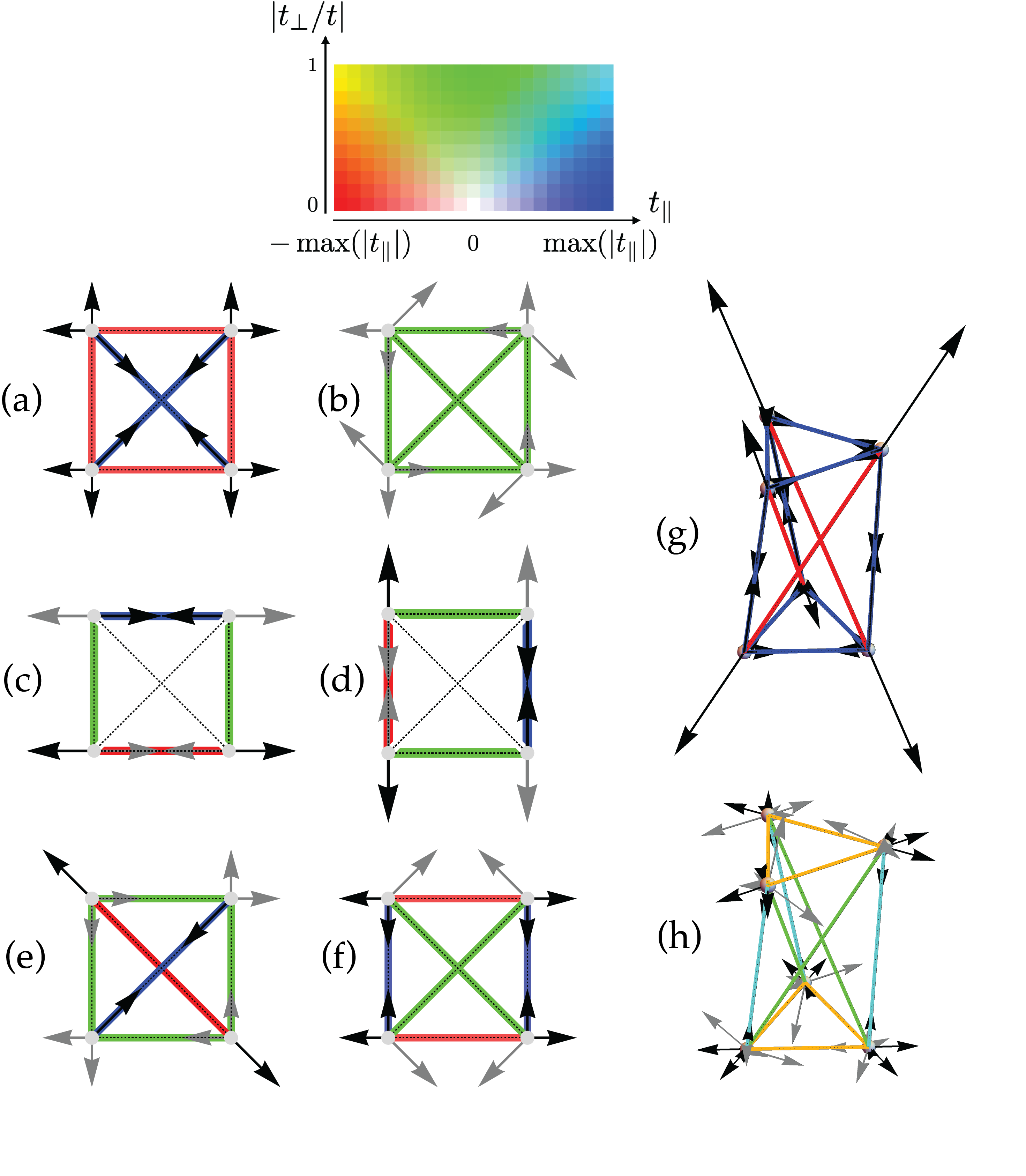}
    \caption{
    Examples of SSSs in prestressed systems.
    (a) A stress-free mechanical network with one SSS.  When the network is prestressed with this SSS, 5 new SSSs arise (b-f).  
    Bonds with $t^{\parallel}, t^{\perp}$ in these SSSs are denoted by the color scheme shown on the top, where $t^{\parallel}$ is denoted by red-white-blue as it goes from negative (compression) to positive (tension), and the relative strength of $t^{\perp}$ over the total $t\equiv \sqrt{(t^{\parallel})^2+(t^{\perp})^2}$ is denoted by green.  The thickness of the bond is proportional to $t$, and the black (gray) arrows denote the component of $t^{\parallel}, (t^{\perp})$ on the nodes.  
    (g) The tensegrity T3-prism has one SSS when it is stress-free. When it is prestressed with this SSS, 20 new SSSs arise, and one example of these SSSs is shown in (h). This example is chosen such that it carries the load of torsion between the top and bottom triangles (see Sec.~\ref{SEC:Projection} for the projection of load onto SSSs).
    }
    \label{FIG:sample-transverse-SSS}
\end{figure}

It might appear confusing how  tension $t$ on a central-force spring can have a component perpendicular to the spring.  This can be understood by realizing that the $\parallel$ and $\perp$ components are defined with respect to the reference configuration, and $t$ denotes an increment of stress in addition to the prestress $t_{p}$. 
In the prestressed reference state, $t_{b,p}$ is along bond $b$ in the reference configuration $\vec{R}_{b,0}$.   In the state after the (infinitesimal) deformation, the total tension is along  bond $b$ in the deformed configuration $\vec{R}_{b}$.  These two bond directions, $\vec{R}_{b,0}$ and $\vec{R}_{b}$, are not parallel in general.  
The $t$ field in  Eq.~\eqref{EQ:Qtot} represents the increment from the pretension to the tension after the deformation, and thus has components parallel and perpendicular to the original bond direction.
In other words, under an (infinitesimal) deformation, both bond (parallel) extension $e^{\parallel}$ and bond rotation $e^{\perp}$ cause increment of stress [Eq.~\eqref{EQ:ES}], and the two components are $t^{\parallel}$ and $t^{\perp}$ respectively.

SSSs in a prestressed system are thus defined as any vectors that satisfy
\begin{align}
\mathbb{Q} 
| t_{\textrm{SSS}} \rangle &=
\mathbb{Q}
\begin{pmatrix}
t^{||}_{\textrm{SSS}}\\
t^{\perp}_{\textrm{SSS}}
\end{pmatrix}
= - | f\rangle
= 0,
\label{EQ:prestressSSS}
\end{align}
which represent respectively parallel and perpendicular change of bond tensions that leave all particles in force balance, for a prestressed network.  In Fig.~\ref{FIG:sample-transverse-SSS} we show two examples of mechanical networks where prestress introduces extra SSSs that involve $t^{\perp}$ components.  This formulation characterizes the tensegrity phenomena~\cite{calladine1978buckminster,connelly1992stability} where prestress generates new SSSs for the system to carry other types of load.

\subsection{Zero modes in prestressed systems}\label{SEC:QCZM}
Prestressed systems have two types of ZMs, type $A$, which  leave $e^{||}=e^{\perp}=0$ on all bonds, and type $B$,  which  violate this relation but are still zero energy modes.  In this section we discuss both types of ZMs.

Type $A$ ZMs in prestressed networks can be defined as modes that live in the null space of $\mathbb{C}$,
\begin{align}
 \mathbb{C} | u_{\textrm{ZM}}^{(A)} \rangle &= 
  \begin{pmatrix}
  e^{||}\\
 e^{\perp}
  \end{pmatrix}
  =| e\rangle =0 .
  \label{EQ:prestressZM}
\end{align}
The number of this type of ZMs
satisfy the Maxwell-Calladine index theorem for a stressed network, which can be proven by applying the rank-nullity theorem on the new $\mathbb{Q},\mathbb{C}$ matrices, 
\begin{align}\label{EQ:prestressedMC}
    N_0^{(A)} - N_S=Nd-N_b d.
\end{align}
Note that the last term is now $N_b d$ for the prestressed network instead of $N_b$ in the stress-free case.  
More rigorously, if we also consider the possibility that not all bonds are stressed, this index theorem should be written as
\begin{align}
    N_0^{(A)} - N_S=Nd-N_b^{\textrm{unstressed}} -N_b^{\textrm{stressed}}d,
\end{align}
where each unstressed bonds  only provide one constraint.

Type $B$ ZMs in prestressed systems are not included in Eq.~\eqref{EQ:prestressZM}.  Instead, they are ZMs due to the fact that some spring constants in $\mathbb{K}$ are negative (from compressively prestressed bonds), and their contribution can cancel out the positive terms.  They are ``fine-tuning'' ZMs which are only zero energy when the prestress satisfies certain conditions so that the positive and negative terms exactly cancel.  This is also the point when varying the magnitude of the prestress starts to cause instability, as we discuss at the end of Sec.~\ref{SEC:Rigidity}. 
One example of such mode is shown in  Fig.~\ref{FIG:rigidity}.  
Interestingly,  because $\mathbb{C}| u_{\textrm{ZM}}^{(B)} \rangle\ne0$ and $\mathbb{Q}\mathbb{K}\mathbb{C}| u_{\textrm{ZM}}^{(B)} \rangle=0$, 
type $B$ ZMs lead to a set of SSSs $| t_{\textrm{SSS}} \rangle=\mathbb{K}\mathbb{C}| u_{\textrm{ZM}}^{(B)}\rangle$.  One trivial example of this type of ZMs is the rigid rotation of the network if the system is under open boundary conditions: the rotation causes $e^{\perp}$ and thus does not satisfy Eq.~\eqref{EQ:prestressZM}, but it is indeed a ZM of the dynamical matrix (this ZM does not require fine tuning of stress but all other type B ZMs do).  A nontrivial example of such fine-tuning ZMs is shown in Fig.~\ref{FIG:rigidity}e where the sum of $e^{\parallel}$ and $e^{\perp}$ terms in the elastic energy  vanish. Type $B$ ZMs are not included in the Maxwell-Calladine index theorem for prestressed systems [Eq.~\eqref{EQ:prestressedMC}].  

Remarkably, although ZMs in prestressed systems are more complicated as we discussed above, the definition of SSSs in Eq.~\eqref{EQ:prestressSSS} is  robust as it relies only on force balance, and is not affected by the positive definiteness of $\mathbb{D}$.  In our discussions below, we mainly focus on these SSSs and show how they form a linear space that efficiently characterizes how load is carried by a prestressed system.

\subsection{Stress response to homogeneous load}\label{SEC:Projection}
One important property of SSSs is that they form a linear space containing \emph{all} possible ways a network can carry stress while keeping all particles in force balance.  
Thus, when the network is under load, actual stress distributions must come from linear combinations of SSSs, and the linear space of SSSs characterizes the capability of a system to carry any external load. 
It is worth pointing out here that we refer to external loads that do not ``introduce new constraints'' to the system, and the meaning of this condition will become clear in the end of Sec.~\ref{SEC:ProjectionDipole}.

Here we first illustrate this formulation using a simple shear on a  stress-free network with all spring constants being 1.
The bond tension in response to this shear can be decomposed in the SSSs linear space because this space form a complete basis for all force-balanced stress distributions, as we discussed above.  The decomposition can be written as
\begin{equation}\label{EQ:Projection}
    |t^{\parallel}\rangle =
    \sum_{i=1}^{N_{\textrm{SSS}}} |t_{\textrm{SSS},i}^{\parallel}\rangle
    \langle t_{\textrm{SSS},i}^{\parallel} | e_{\textrm{affine}}^{\parallel} \rangle ,
\end{equation}
where $e_{\textrm{affine}}^{\parallel}$ is the bond extension if the strain were affine (i.e., homogeneous shear strain).  
The sum runs in the $N_{\textrm{SSS}}$ dimensional linear space of all SSSs,  $|t_{\textrm{SSS},i}^{\parallel}\rangle $. 
This relation is  straightforward to prove as follows: when 
a strain is imposed on the system 
%which would cause bond extension $e_{\textrm{affine}}^{\parallel}$ 
%is externally imposed 
(e.g., via Lees-Edwards boundary conditions in a computational model, or in the bulk of a mechanical network strained from the boundary), in addition to affine bond extensions $e_{\textrm{affine}}^{\parallel}$, the system responds by particle displacements $u$ (e.g., nonaffine deformations) to minimize the elastic energy while maintaining this macroscopic strain, so the resulting bond extensions are 
\begin{equation}\label{EQ:eparaANA}
    |e^{\parallel}\rangle
    = | e_{\textrm{affine}}^{\parallel} \rangle + \mathbb{C}^{\parallel} |u\rangle .
\end{equation}
If all spring constants are 1, we have $|t^{\parallel}\rangle=|e^{\parallel}\rangle$.
As we mentioned above, $|t^{\parallel}\rangle$ must belong to the SSSs linear space, 
and the linear combination coefficients 
%of $t^{\parallel}$ to the SSSs 
are
\begin{equation}
    c_i = \langle t_{\textrm{SSS},i}^{\parallel} | t^{\parallel}\rangle 
   % = | e_{\textrm{imposed}}^{\parallel} \rangle + \mathbb{C} |u\rangle 
   = \langle t_{\textrm{SSS},i}^{\parallel} | e_{\textrm{affine}}^{\parallel} \rangle
   +\langle t_{\textrm{SSS},i}^{\parallel} |\mathbb{C}^{\parallel} |u\rangle 
   .
\end{equation}
The second term has to vanish because $\langle t_{\textrm{SSS},i}^{\parallel} |\mathbb{C}^{\parallel}=(\mathbb{Q}^{\parallel}|t_{\textrm{SSS},i}^{\parallel}\rangle)^T=0$.  This proves the decomposition in Eq.~\eqref{EQ:Projection}.

Three comments can be made from this result. First, if the shear strain has no overlap with any SSSs in the system,  $\langle t_{\textrm{SSS},i}^{\parallel} | e_{\textrm{affine}}^{\parallel}\rangle =0$ for all $i$, the system can not carry the load. What would happen physically is that the system yields until a new SSS, able to carry that load, emerges.  
Second, the same approach can also be applied to other types of load, such as a hydrostatic pressure. It is important to note here that any component in the imposed strain that can be written in terms of $\mathbb{C}^{\parallel} |u\rangle $ will not cause stress---it corresponds to strain that will be relaxed by degrees of freedom available to the system.  
Third, a similar formulation can also be developed with loads applied via specifically controlled boundary displacements, where SSSs are defined as bond tensions leaving internal sites in force balance, and the strain $e$ can be applied from  bonds connected to boundaries~\cite{zhang2018fracturing}.

The relation is slightly more complicated when the spring constants are different for each spring,
\begin{equation}\label{EQ:ProjectionK}
    |t^{\parallel}\rangle =
    \sum_{i,j}^{N_{\textrm{SSS}}} |t_{\textrm{SSS},i}^{\parallel}\rangle
    \lbrack ((\mathbb{K^{\parallel}})^{-1})_{ss} \rbrack^{-1}_{ij}
    \langle t_{\textrm{SSS},j}^{\parallel} | e_{\textrm{affine}}^{\parallel} \rangle ,
\end{equation}
where  $((\mathbb{K^{\parallel}})^{-1})_{ss}$ is the inverse of the spring constant matrix projected to the SSS linear space (see detailed definition in App.~\ref{APP:Projection}).

This relation can readily be generalized to the pre-stressed case, where
\begin{equation}\label{EQ:ProjectionKS}
    |t\rangle =
    \sum_{i,j}^{N_{\textrm{SSS}}} |t_{\textrm{SSS},i}\rangle
    \lbrack (\mathbb{K}^{-1})_{ss} \rbrack^{-1}
    \langle t_{\textrm{SSS},j} | e_{\textrm{affine}} \rangle .
\end{equation}
%where $\mathbb{K}$ includes both 
A detailed proof of this relation is included in App.~\ref{APP:Projection}.  It is worth noting that the response $|t\rangle$ calculated here is the addition to the prestress in response to the load, so the total stress in the system is $|t_p \rangle +|t\rangle$ where $t_p$ only has longitudinal components and $t$ has both longitudinal and transverse components (with respect to the bond direction in the prestressed reference state).

\subsection{Stress response to force dipoles}\label{SEC:ProjectionDipole}
Similar projections can also be applied to compute the stress response of a prestressed system to local forces, such as a force dipole acting on a bond. 

\begin{figure}[H]
    \centering
    \includegraphics[width=0.4\textwidth]{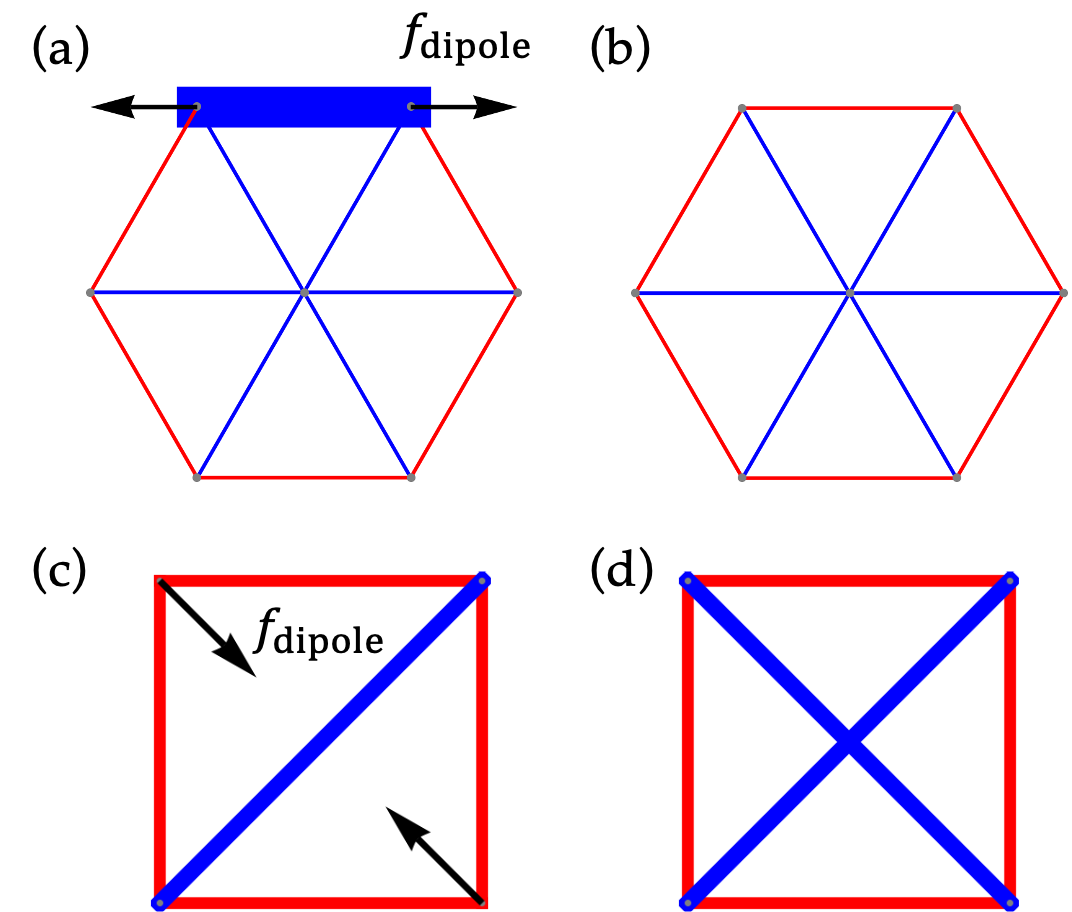}
    \caption{Stress response to dipole forces. (a) Bond stress response of a mechanical network to a force dipole $f_{\textrm{dipole}}$ (black arrows) acting on a bond in the network. Blue and red marks the tension $t_{\textrm{rsp}}>0$ and compression $t_{\textrm{rsp}}<0$ on the bonds in response to the force dipole, with the thickness of the bond proportional to the magnitude of the tension/compression. (b) The sum of the external dipole and the network response, $t_{\textrm{dipole}}+t_{\textrm{rsp}}$, is a SSS of the network.  (c) Bond stress response of a mechanical network to a force dipole acting on a pair of particles not connected by a bond.  This force dipole can be viewed as an additional bond in the network, such that $t_{\textrm{dipole}}+t_{\textrm{rsp}}$ is a SSS of the network with the added bond (d).  The original structure in (c) does not have any SSS. 
    }\label{FIG:dipole}
\end{figure}

The main difference between this case of a force dipole and the external strain case discussed in Sec.~\ref{SEC:Projection} is that 
the force-balance condition here needs to include the external forces, 
%involves the balance between the external force and the network response,
\begin{align}\label{EQ:FBDipole}
    \mathbb{Q}|t_{\textrm{rsp}}\rangle = -|f_{\textrm{rsp}} \rangle = |f_{\textrm{external}}\rangle
\end{align}
where $f_{\textrm{external}}$ is the externally imposed force %on the particles, 
and $t_{\textrm{rsp}}$ is the tension response of the system, which causes $f_{\textrm{rsp}}$. $f_{\textrm{rsp}}$ cancels with $f_{\textrm{external}}$, leaving the system in force balance.

When the external forces take the form of a force dipole on a bond $b$, they can be written as 
\begin{align}\label{EQ:FDipole}
    |f_{\textrm{external}}\rangle
    = |f_{\textrm{dipole}}\rangle
    = - \mathbb{Q} |t_{\textrm{dipole}}\rangle 
    = - \mathbb{Q}\mathbb{K} |b\rangle 
\end{align}
where $|b\rangle $ is a vector in the bond space with 1 on the bond where the dipole is imposed, and 0 on all other bonds.  This sets the magnitude of the dipole force to be $k_b\cdot 1$, the value of which is irrelevant for the linear theory.  The sign convention of this equation follows from Eq.~\eqref{EQ:Qtot} where $t_{\textrm{dipole}}$ causes $f_{\textrm{dipole}}$. 
Combining Eqs.~\eqref{EQ:FBDipole} and \eqref{EQ:FDipole} we have
\begin{align}
    \mathbb{Q} (|t_{\textrm{dipole}}\rangle
    +|t_{\textrm{rsp}}\rangle)=0,
\end{align}
indicating that the total tension field  $t_{\textrm{dipole}}+t_{\textrm{rsp}}$ (sum of the external dipole and the network response) is a SSSs of the network.  The response can then be derived using a similar method as discussed in Sec.~\ref{SEC:Projection}, and we have (as detailed in App.~\ref{APP:Dipole})
\begin{equation}\label{EQ:ProjectionDipole}
    |t_{\textrm{rsp}}\rangle =
    - |t_{\textrm{dipole}}\rangle +
    \sum_{i,j}^{N_{\textrm{SSS}}} |t_{\textrm{SSS},i}\rangle
    \lbrack (\mathbb{K}^{-1})_{ss} \rbrack^{-1}_{ij}
    \langle t_{\textrm{SSS},j} | b \rangle 
    .
\end{equation}
In the simple case of $\mathbb{K}=k\mathbb{I}$ (all springs having the same spring constant which only applies to stress-free networks), the second term on the right hand side of the equation reduces to  the ``quasi-localized SSSs'' defined in the stress-free case, characterizing dipole responses~\cite{sussman2016spatial,lerner2018quasilocalized}. Here we extend this concept to prestressed networks, and our formulation can be used to not only compute dipoles with force along the bond, but also ``transverse'' or ``tilted'' dipoles where the force pair is perpendicular or in an arbitrary angle to the bond (see example in Fig.~\ref{FIG:dipole}).

A ``dipole stiffness'' $\kappa$ can then be defined to characterize this response
\begin{equation}
\kappa \equiv 
\frac{\langle{f_{\textrm{dipole}}|f_{\textrm{dipole}}}\rangle}{\langle{f_{\textrm{dipole}}|u_{\textrm{rsp}} }\rangle},
\label{EQ:DipoleK}
\end{equation}
where $\langle{f_{\textrm{dipole}}}|$ serves the purpose of projecting the force and displacements to the direction of the external force dipole.  
{A different version of (longitudinal only) dipole stiffness has been introduced in Ref.~\cite{rainone2020statistical} where $u_{\textrm{rsp}}$ of the whole system is used instead of its projection to the dipole was used. 
In contrast, the dipole stiffness defined here treats the whole system as a black box}, and only extracts the stiffness from the force-distance relation between the two particles where the dipole acts on.  The simplicity of this definition allows it to be directly implemented in experiments via methods such as microrheology.

Using the SSSs linear space, the dipole stiffness to a local force dipole on bond $b$ is thus given by (details in  App.~\ref{APP:Dipole}):
\begin{equation}\label{EQ:DipoleStiffness}
    \kappa_{b}
    = \frac{2k_b^2}{ k_b -  \displaystyle \sum_{i,j}^{N_{\textrm{SSS}}} \langle b |t_{\textrm{SSS},i}\rangle
    \lbrack (\mathbb{K}^{-1})_{ss} \rbrack^{-1}
    \langle t_{\textrm{SSS},j} | b \rangle},
\end{equation}
where $k_b$ is the spring stiffness for bond $b$. 

Another interesting case of force dipole is when the dipole acts on a  pair of particles that are not connected in the network (which can be called ``non-local dipoles'').  The analysis of force-balance in this case can be done by considering the external force dipole as \emph{introducing an additional constraint into the network}.  Thus, the stress distribution is a SSS of the network with this additional ``auxiliary bond'', instead of the original network.  
In App.~\ref{APP:Dipole} we also derive the dipole stiffness of this case.  Examples of both types of dipoles are shown in Fig.~\ref{FIG:dipole}.
%, and Fig.~\ref{FIG:Trig2} illustrates an example of dipoles on a pair of particles that are not connected.

%%%%%%%%%%%%%%%%%%%%%%%%%%%%%%%%%%%%%
%%%%%%%%%%%%%%%%%%%%%%%%%%%%%%%%%%%%%
%%%%%%%%%%%%%%%%%%%%%%%%%%%%%%%%%%%%%
\section{Prestressed elasticity of triangular lattices}\label{SEC:Triangular}
%%%%%%%%%%%%%%%%%%%%%%%%%%%%%%%%%%%%%
%%%%%%%%%%%%%%%%%%%%%%%%%%%%%%%%%%%%%
%%%%%%%%%%%%%%%%%%%%%%%%%%%%%%%%%%%%%

In this section we use prestressed triangular lattices to illustrate the mathematical tools for prestressed elasticity  we discussed in Secs.~\ref{SEC:Prestress} and \ref{SEC:SSS}. Models of prestressed triangular lattices have been introduced in the literature of force network ensembles, as an example system where multiple internal stress distributions are  compatible with given boundary stress~\cite{Snoeijer2004,Tighe2008,tighe2010force}.  Here we discuss how our SSSs formulation can be applied to these lattices to both systematically generate all possible prestress distributions, and solve for vibrational modes and stress response to external load, revealing interesting prestress effects.

\begin{figure*}%[H]
    \centering
    \includegraphics[width=0.9\textwidth]{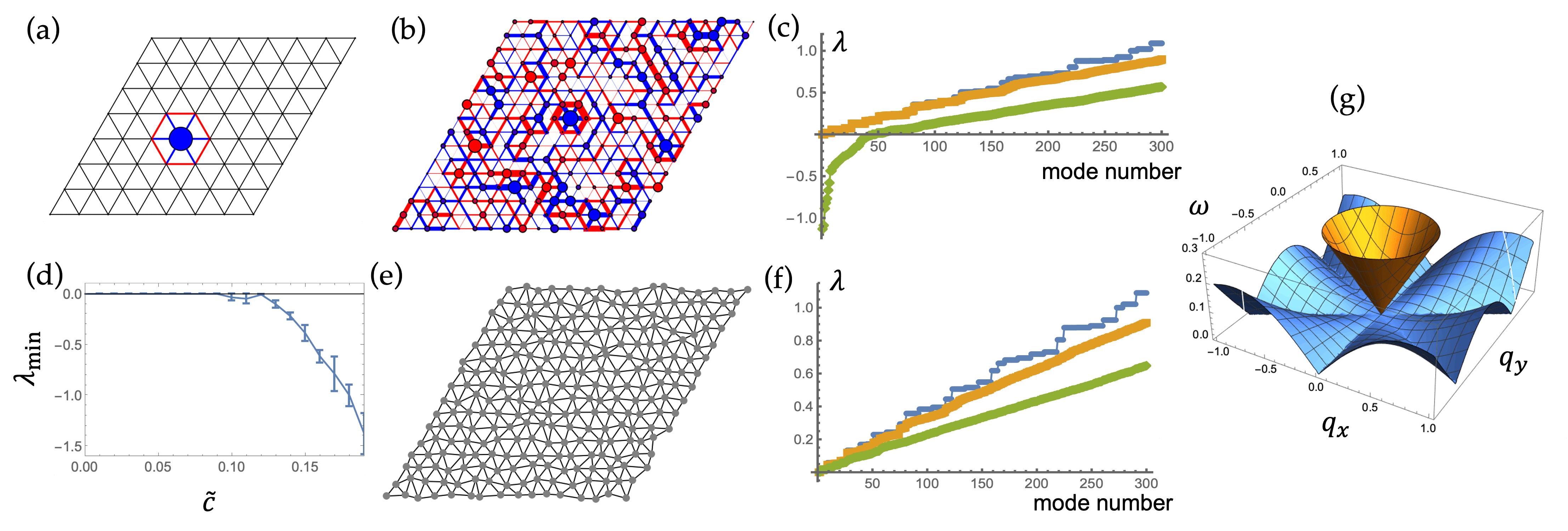}
    \caption{Vibrational modes of prestressed triangular lattices.  (a) One site-localized SSS $\vert t_{\ell}\rangle$ on a triangular lattice, where bonds surrounding the hexagon carry compression, and bonds in the hexagon carry tension. The size of the blue disk on the center site represents the strength of this SSS. (b) A prestressed triangular lattice with the prestress $t_p$ generated using a linear combination of site-localized SSSs with independent coefficients on each site, drawing from a Gaussian distribution with mean 0 and standard deviation $\tilde{c}=0.12$.
    (c) Lowest 300 eigenvalues $\lambda$ of the dynamical matrix of prestressed triangular lattices of size $32\times 32$, at $\tilde{c}=0$ (blue), $\tilde{c}=0.1$ (orange), $\tilde{c}=0.2$ (green), from top to bottom.
    (d) The lowest eigenvalue $\lambda_{\textrm{min}}$ as a function of $\tilde{c}$. 
    (e) A triangular lattice with positional disorder where the random displacements of the nodes have standard deviation $\tilde{u}=0.12$. 
    (f) Lowest 300 eigenvalues $\lambda$ of the dynamical matrix of a $32\times 32$ triangular lattices with positional disorder, at $\tilde{u}=0$ (blue), $\tilde{u}=0.1$ (orange), $\tilde{u}=0.2$ (green), from top to bottom. (g) Phonon disperson relation of the lowest band of triangular lattices with no prestress (yellow, upper), and critical compressive prestress (blue, lower) where modes along $\Gamma M$ approach instability.
    }
    \label{FIG:Trig1}
\end{figure*}

In the stress-free case, triangular lattices have a coordination number $z=6>2d$ and thus the number of SSSs per site is $\frac{z}{2}-2=1$.  In fact, a localized SSS $\vert t_{\ell}\rangle$ can be defined around each site $\ell$ as shown in Fig.~\ref{FIG:Trig1}a.  These site-localized SSSs, together with the homogeneously state of self stress (all bonds carrying the same tension),  form a complete basis to decompose any SSSs on the triangular lattice.

\begin{figure*}%[H]
    \centering
    \includegraphics[width=0.85\textwidth]{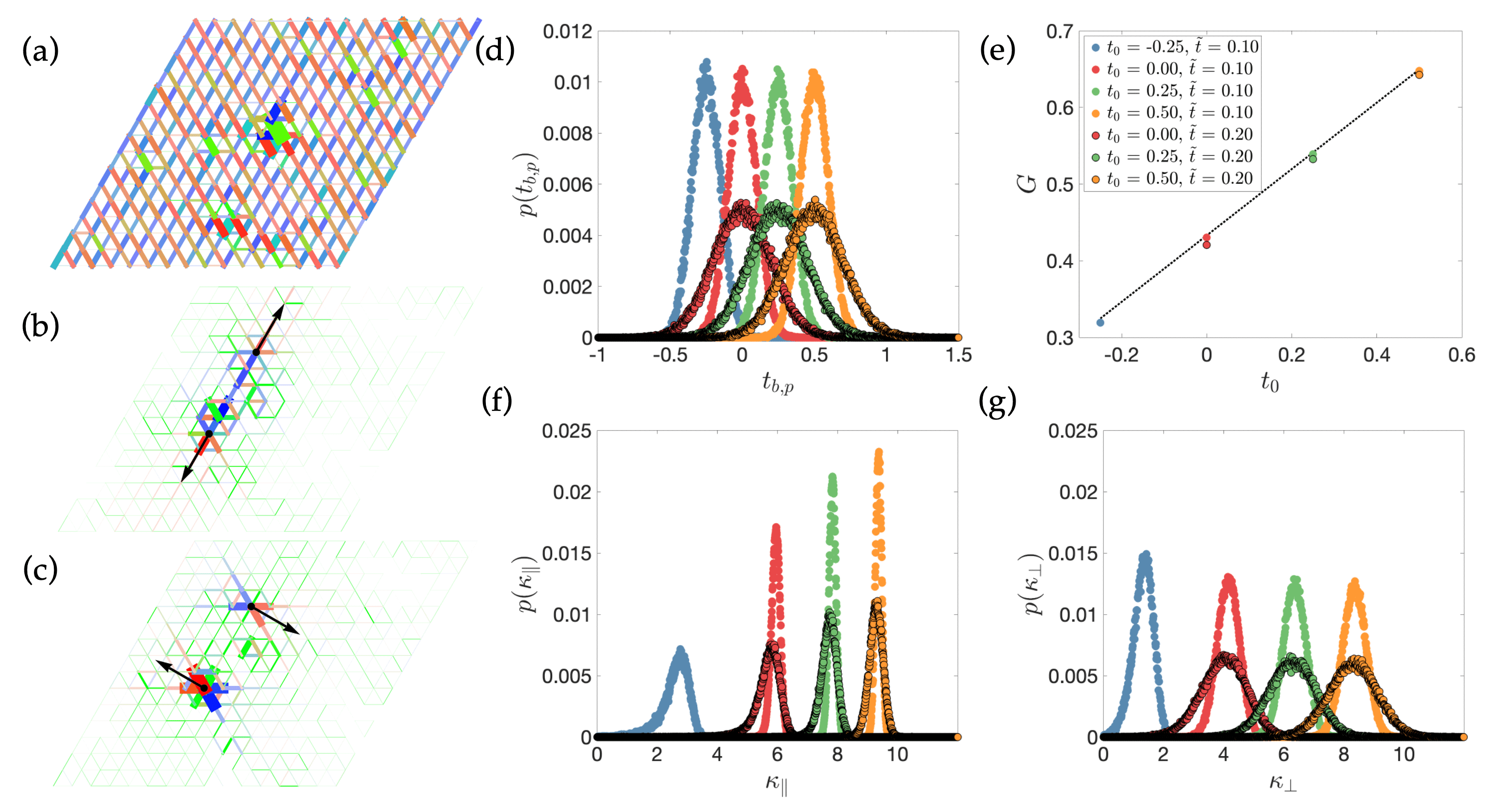}
    \caption{Prestressed triangular lattices under load.  (a-c) An example of a prestressed triangular lattice (the one shown in Fig.~\ref{FIG:Trig1}b) under shear load $\epsilon_{xy}$ (a), a longitudinal dipole (b), and a transverse dipole (c), where the force dipoles are shown as black arrows.  The color scheme of bond tensions ($t^{\parallel}$ and $t^{\perp}$) is the same as in Fig.~\ref{FIG:sample-transverse-SSS}.  
    (d) Probability distributions of pretensions on bonds of the lattices we tested for load bearing, with different mean $t_0$ and standard deviation $\tilde{t}$ of pretension on bonds [legend shown as inset of  (e)]. (e) Shear modulus $G$ of these lattices as a function of mean pretension $t_0$. (f-g) Probability distributions of the parallel (f) and transverse (g) local dipole stiffness of these lattices. $20\times 20$ lattices are used in the computations in (d-g), and we average over 100 realizations of disorder for each set of parameters.
    }
    \label{FIG:Trig2}
\end{figure*}

In particular, we can generate an ensemble of prestress $\vert t_{p}\rangle$ on triangular lattices by taking an arbitrary coefficient $c_{\ell}$ for each site-localized SSS $\vert t_{\ell}\rangle$ and sum them up.  In Fig~\ref{FIG:Trig1}b we show an example of such a prestressed state. 

These prestressed lattices can be physically constructed by choosing the rest length of each bond $b$ such that when they are at the length in the regular triangular lattice, the tension/compression they carry is exactly the value $t_{b,p}$ in the prescribed SSSs.  Because the total force is balanced at each site for any SSS, all bonds will stay at the length in the regular lattice, and thus we obtain a triangular lattice with \emph{regular geometry and a prescribed SSS}.

What are the mechanical properties of such prestressed triangular lattices? Here we study them from two perspectives: vibrational modes and load-bearing capabilities.  The vibrational modes can be calculated based on the quadratic expansion in Eq.~\eqref{EQ:dVb} for each bond, which leads to the dynamical matrix defined in Eq.~\eqref{EQ:DM}.

In Fig.~\ref{FIG:Trig1}c we show eigenvalues of the dynamical matrix of the triangular lattice at three levels of prestress (by assigning $c_{\ell}$ from a Gaussian distribution with mean at $0$ and standard deviation at $\tilde{c}$).  Prestress significantly affects the vibrational modes especially at low frequencies.  In particular, negative eigenvalues start to appear when the fluctuations of the SSSs goes beyond a critical level,  $\tilde{c}>\tilde{c}^*$ (Fig.~\ref{FIG:Trig1}c,d).  This indicates that by increasing the fluctuation of prestress (where the mean remains 0), unstable modes appear, although the system remains in force balance.  This is in alignment with our discussion on prestressed rigidity and type $B$ ZMs of prestressed systems in Sec.~\ref{SEC:Rigidity}.

To contrast this result, we also consider triangular lattices with positional disorder but no prestress (Fig.~\ref{FIG:Trig1}e).  In this case, we move site $\ell$ by a random displacement $\vec{u}_{\ell}$, the $x,y$ components of which being generated  from a Gaussian distribution with mean at $0$ and standard deviation at $\tilde{u}$.  We also plot the eigenvalues of the dynamical matrix of this case in Fig.~\ref{FIG:Trig1}f.  Although positional disorder also affects the eigenvalues, it mostly smooths out the regular lattice eigenmodes, and does not lead to qualitative change at low frequencies and does not generate unstable modes.  Note that to make a fair comparison we chose the same values for $\tilde{c}$ and $\tilde{u}$, which represents the same level of disorder, under the simple convention we took where the bond length and spring constant on the triangular lattice both being unity.

The load-bearing capabilities of prestressed triangular lattices can be analyzed by the SSSs formalism discussed in Sec.~\ref{SEC:SSS}.  At any given realization of disordered prestress, we can compute the linear space of SSSs, and use Eq.~\eqref{EQ:ProjectionKS} and \eqref{EQ:ProjectionDipole} to find the stress response of the system to external load (Fig.~\ref{FIG:Trig2}a-c).  
In particular, to characterize the homogeneous component of prestress (hydrostatic pressure), we also introduce a constant pretension $t_0$ on all bonds, in addition to the fluctuations from $|t_{\ell}\rangle$.  As a result, the tension on each bond obeys a normal distribution with mean at $t_0$ and standard deviation at $\tilde{t}=2\tilde{c}$ (as each bond appears in 4 site-localized SSSs around it), as shown in Fig.~\ref{FIG:Trig2}d.  
We apply  two types of load, simple shear and dipole forces, on these prestressed triangular lattices, 
and the results are shown in Fig.~\ref{FIG:Trig2}.  
This analysis shows that for both shear modulus and stiffness against dipole forces, their mean is mainly controlled by $t_0$ (where compression decreases stiffness and tension increases stiffness) whereas $\tilde{t}$ slightly decreases these stiffness, as well as broadens them.  In addition, the stress response demonstrates interesting heterogeneities due to the disordered prestress, an effect we will discuss more in Sec.~\ref{SEC:Amorphous}.

Effects of prestress on mechanics can also be seen from the perspective of phonon structures of regular lattices (Figs.~\ref{FIG:rigidity}c and 
\ref{FIG:Trig1}g).    
It is straightforward to see that negative prestress (compression) destabilizes the originally stable triangular lattice (Fig.~\ref{FIG:Trig1}g), and positive prestress (tension) stabilizes the honeycomb lattice (Fig.~\ref{FIG:rigidity}c), which was originally unstable against shear.  Interestingly, the modes that first become unstable in the triangular lattice as negative prestress increase are the modes along the $\Gamma M$ direction in the first Brillouin zone, indicating that the type of modes that first become unstable are the modes that zig-zag between the straight lines of bonds, agreeing with recent studies of strain-localization in lattice models~\cite{BORDIGA2021104198}.

%%%%%%%%%%%%%%%%%%%%%%%%%%%%%
%%%%%%%%%%%%%%%%%%%%%%%%%%%%%
%%%%%%%%%%%%%%%%%%%%%%%%%%%%%
%%%%%%%%%%%%%%%%%%%%%%%%%%%%%
\section{Prestressed elasticity of amorphous solids of soft repulsive particles}\label{SEC:Amorphous}
%%%%%%%%%%%%%%%%%%%%%%%%%%%%%
%%%%%%%%%%%%%%%%%%%%%%%%%%%%%
%%%%%%%%%%%%%%%%%%%%%%%%%%%%%
The new $\mathbb{Q},\mathbb{C}$ matrix approach described in Sec.~\ref{SEC:SSS} provides a set of tools to analyze stress-bearing capabilities and mode structures of prestressed systems, well beyond the triangular lattices just described. Here we apply this method to a computational model for amorphous solids composed of soft repulsive particles, where, akin to real materials, prestresses are naturally introduced by the preparation (solidification) protocol. Given the preparation protocol chosen for this study, the prestress is the result of 
%, on which this method can be applied. 
%Such amorphous solids feature prestress both from c
both compression and frozen-in structural disorder. 

%We present results from applying the new $\mathbb{Q},\mathbb{C}$ matrix methods to this model system.

\subsection{3D numerical simulations of amorphous solids of soft repulsive particles}\label{SEC:Model}
The numerical model describes a suspension of particles with soft repulsive interactions given by a truncated and shifted Lennard-Jones potential~\cite{weeks1971role}. The potential energy, for a pair of two
particles $\ell$ and $\ell'$ separated by a center-to-center distance $R_{b}\equiv |\vec{R}_\ell - \vec{R}_\ell'|$, is 
$V(R_{b}) = 4\epsilon[(a_{b}/R_{b})^{12} - (a_{b}/R_{b})^6] + \epsilon$ for particle pairs closer than the cutoff distance $R_{b} \leq r_c\equiv 2^{1/6}a_{b}$, otherwise $V(R_{b}) = 0$, following the notation we defined in Sec.~\ref{SEC:Prestress}. Here $a_{b} =(a_{\ell}+a_{\ell'})/2$ with $a_{\ell}$ and $a_{\ell}$ being the diameters.  
%, and $R_b\equiv |\vec{R}_b|$ is the center-to-center distance between the two particles, 
The size poly-dispersity is introduced by drawing the diameter of each particle $a_i$ from a Gaussian distribution with mean $a$ and variance of $10\%$. The parameter $\epsilon$ is the unit energy and $a$ is the unit length in the simulations.  Particles closer than the cutoff distance (beyond which the inter-particle force vanishes), $R_b\le r_c$, are considered in contact.  
All simulations used here have volume fraction $\phi \approx 70\%$ and consist of $10^4 (10976)$ particles in a cubic box of linear size $L=20.36$, unless otherwise specified. The initial samples are prepared by first melting face-centered-cubic (FCC) crystals of particle at volume fraction $0.7$ at $T = 5.0 \epsilon/k_B$. 
We then cool the particle assemblies down to a temperature $0.001 \epsilon/k_B$ through an NVT molecular dynamics (MD) protocol as described in Ref.~\cite{vasisht2020computational}. The cooling rate $\Gamma$ varies from $5\times 10^{-2}$ to $5\times 10^{-6}~\epsilon/(k_B\tau_0)$, where $\tau_0 = a\sqrt{m/\epsilon}$ is the MD time unit with $m$ the particle mass. Subsequently, each sample is brought to the closest local energy minimum using a conjugate gradient (CG) algorithm. The configurations prepared in this way are amorphous solids, as we verify by measuring their viscoelastic response \cite{vasisht2020computational}, whose properties depend on the cooling rate utilized. These solids feature prestress that comes both from the homogeneous compression %due to the non-negligible particle overlaps at this high particle volume fraction 
and from the structural disorder due to the size polydispersity. 

From all contacting particle pairs, we compute the virial stress tensor 
\cite{irving1950statistical} (neglecting the kinetic term):
\begin{equation}\label{EQ:Virial}
   \displaystyle \sigma_{ij} = \frac{1}{V}\sum_{\ell}\sum_{\ell'>\ell}R_{b,i}t_{b,j} ,
\end{equation}
where $V$ is the volume of the system, $R_{b}$ is the distance between particle $\ell$ and $\ell'$ (length of bond $b$), and $t_{b}$ is the force on bond $b$.

\begin{figure*}%[H]
    \centering
    \includegraphics[width=\textwidth]{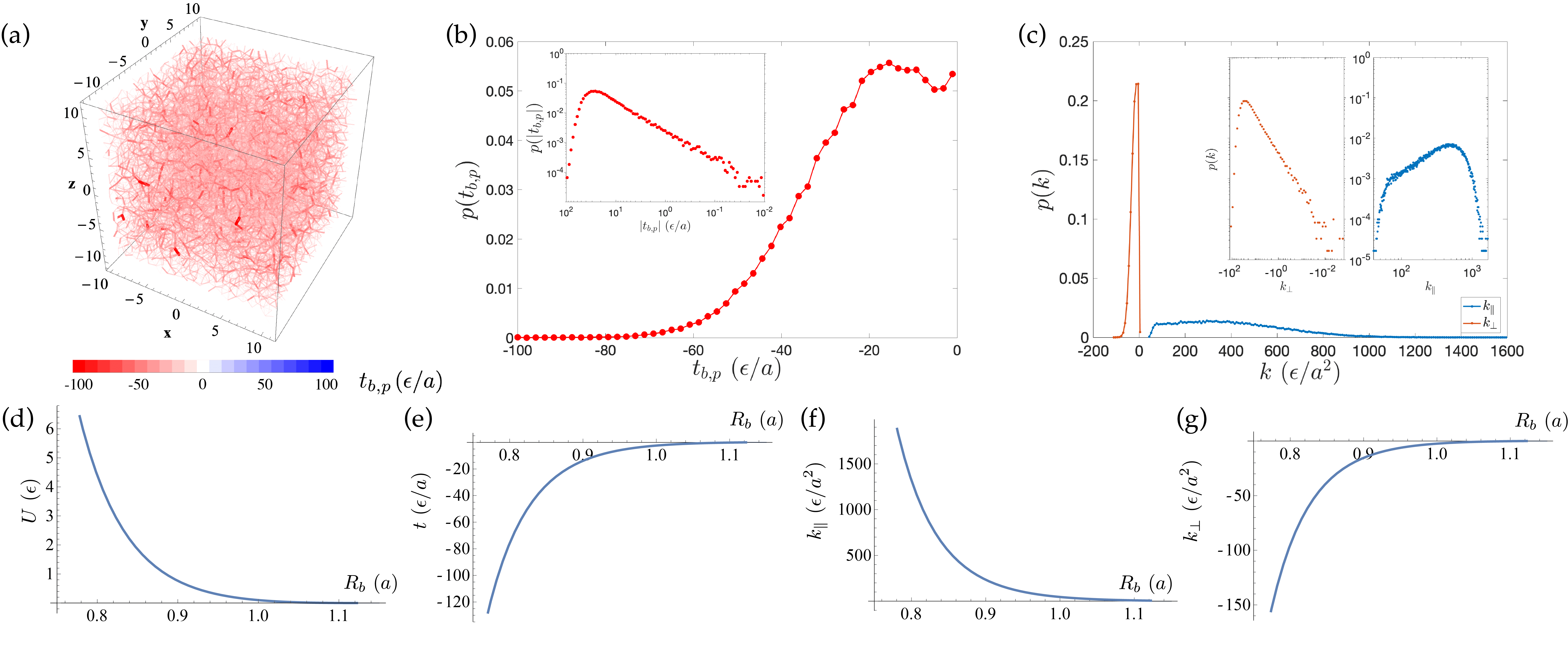}
    \caption{Prestress in  amorphous solids of repulsive particles. (a) Pretension on bonds $t_{b,p}\equiv \frac{dU(r_{ij})}{dr_{ij}}$ where $b$ is the bond connecting particles $i,j$.
    The system size is $N=10976$ with preparation cooling rate $\Gamma = 5\times 10^{-6}~\epsilon/(k_B\tau_0)$. (b-c) Distributions of $t_{b,p}$ (b), and spring constants $k_\parallel \equiv k_b$ and $k_\perp \equiv \displaystyle \frac{t_{b,p}}{ |\vec{R}_{b,0}|}$ (c), for the configuration shown in (a).
    (d-g) $U$, $t$, $k_\parallel$, and $k_\perp$ for the pairwise potential used in the model as functions of distance $R_b$.
    }
    \label{FIG:PrestressVisual}
\end{figure*}

In Fig.~\ref{FIG:PrestressVisual} we show a typical configuration where prestress on the bonds 
are visualized.  These $t_{b,p}$ terms are negative as the system is compressed.  
For each cooling rate, we prepare 5 statistically independent samples. All quantities investigated here have been averaged over this set of samples and we obtain error bars from sample to sample fluctuations.

We also examine these samples as they are sheared using Lees-Edwards boundary conditions and a shear 
rate $\dot{\gamma}$, by solving Newton equations of motions with a drag force that guarantees minimal inertia effects as discussed in ~\cite{vasisht2020computational}. All simulations are performed with LAMMPS~\cite{plimpton1995fast}, suitably modified to include the particle size polydispersity and the interactions discussed above, and using a shear rate $\dot{\gamma} = 10^{-4} \tau_0^{-1}$. 

With the bond network and particle coordinates, we build the compatibility and equilibrium matrices $\mathbb{Q}$ and $\mathbb{C}$, to study the prestressed elasticity using the general formalism described in the previous sections. In this 3D network, each bond has one longitudinal $t^{||}$ and two transverse directions $t^{\perp}$ for tension increments.  
We take the convention that the first transverse direction  $\hat{R}_{b,0}^{\perp,1} \equiv \hat{R}_{b,0} \times \hat{z} / |\hat{R}_{b,0} \times \hat{z}|$ is the unit vector of the cross product between the bond unit vector $\hat{R}_{b,0}$ and the $z$-axis $(0,0,1)$, and the second transverse direction  $\hat{R}_{b,0}^{\perp,2} \equiv \hat{R}_{b,0} \times \hat{R}_{b,0}^{\perp,1} $ is the unit vector of the cross product between the bond vector and the first transverse direction [the subscript $0$ signifies the reference (undeformed but prestressed) state]. 

Once the $\mathbb{Q}$ and $\mathbb{C}$ matrices are constructed, we solve their null space and obtain ZMs and SSSs of the computational model.  The samples we studied are in general deep in the  solid phase and  exhibit no ZMs besides the 3 trivial translations.  
When solving for SSSs in large and dense systems as the samples we generated, the null space for the equilibrium matrix $\mathbb{Q}$ may have very high dimensions. We used the SPQR\_RANK package~\cite{foster2013algorithm} from SuiteSparse~\cite{suitesparse}, which is a high performance sparse QR decomposition package and can provide a reliable determination of null space basis vectors for large sparse matrices, to solve for SSSs efficiently and reliably, especially when there are lots of degeneracies for diagonalizing $\mathbb{Q}$. Similarly, SPQR\_RANK is also suitable when solving for the null space of $\mathbb{Q}^T$ to get ZMs in large systems.

%%%%%%%%%%%%%%%%%%%%%%%%%%%%%
%%%%%%%%%%%%%%%%%%%%%%%%%%%%%
%%%%%%%%%%%%%%%%%%%%%%%%%%%%%
%%%%%%%%%%%%%%%%%%%%%%%%%%%%%

\subsection{Calculating stress response to shear using states of self-stress}
We begin by demonstrating the power of the SSSs as ``stress-eigenstates'' that characterize the stress-bearing capabilities of a system.  To this end, we compute all SSSs as described above, and use Eq.~\eqref{EQ:ProjectionKS} to calculate the change of tension, $t^{\parallel}$ and $t^{\perp}$, as the system is under macroscopic shear, which we call ``prestressed response'' here, as this is the difference between the stress under load and the prestress.

To contrast this result, we also calculated the tension increment treating the system as a stress-free network, i.e., as characterized by equilibrium matrix $\mathbb{Q}^{\parallel}$ instead of $\mathbb{Q}$  (where the geometry of the contact network is kept the same).  The resulting stress increment (which only has the $t^{\parallel}$ component) is then calculated using Eq.~\eqref{EQ:ProjectionK}, and we will call this the  ``stress-free response''.
At the same time, we also measure the change of the stress in the system in a numerical experiment where the system is under a small shear deformation imposed  quasistatic or at finite rate (as specified in captions of Figs.~\ref{Fig:visual_tension_inc} and \ref{Fig:tension_inc_predict}), and we call this the ``actual response''.

In the next two subsections we compare these different results on tension increment in terms of both spatial heterogeneity and shear modulus.

\subsection{Spatial heterogeneity of stress response}
\begin{figure*}%[H]
    \centering
    \includegraphics[ width=0.75\textwidth]{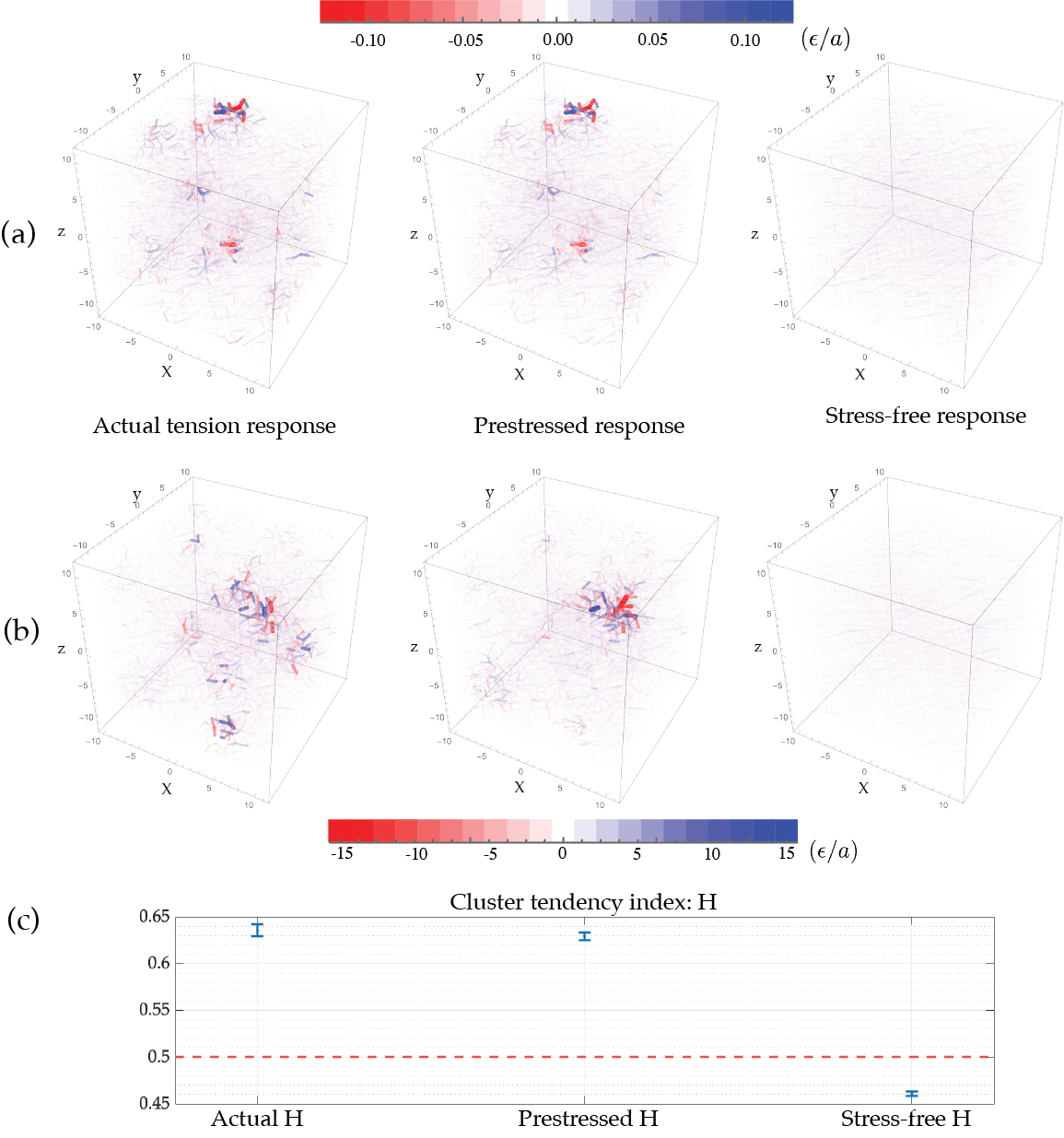}
    \caption{Spatial heterogeneity of tension increment of prestressed networks under shear strain.  We compare the actual tension response $t$ measured in simulation to the tension increment calculated with and without prestress, for  a quasistatic shear deformation at strain 0.01\% in (a) and a finite rate shear deformation at strain 1\% in (b) (system size  $N=10976$ and cooling rate $\Gamma = 5\times 10^{-2}~\epsilon/(k_B\tau_0)$).  
    (c) Clustering tendency index $H$ averaged over three different samples calculated using stress-free and prestressed elasticity compared to the actual $H$ measured at quasistatic shear strain $0.01\%$.  
    The red dashed line indicates the threshold to determine whether or not the tension change is clustered.
    }
    \label{Fig:visual_tension_inc}
\end{figure*}
One significant effect of prestress on the mechanical response of amorphous solids is the spatial heterogeneity of the stress increment, 
in contrast to the stress-free response (where the geometry is kept the same). 
Fig.~\ref{Fig:visual_tension_inc} shows a visualization of the tension increment in the model amorphous solids we study here.

Remarkably, this heterogeneity is accurately depicted from the SSSs calculation when prestress is included. 
The tension increment is computed following Eq.~\eqref{EQ:ProjectionKS}. The resulting $t$ field includes both $t^{||}$ and $t^{\perp}$, although only $t^{||}$ is shown.  
To be precise, $t^{\perp}$ manifests as rotations of the bonds in the real response (which causes the total tension to change direction). 
In contrast, the calculated stress response of the system ignoring the prestress effect, i.e., using Eq.~\eqref{EQ:ProjectionK}, is almost completely homogeneous. Thus, in these dense systems, \emph{the prestress controls the spatial heterogeneity of the mechanical response of the system.} Treating the system as ``stress-free'' misses important signatures of heterogeneities.

We characterize the spatial heterogeneity of the tension increment using a clustering tendency index: the Hopkins statistic $H$~\cite{hopkins1954new}, with a value close to 1 indicating the data is highly clustered, and a value around 0.5 from random data.
Fig.~\ref{Fig:visual_tension_inc}(c) shows that the prestressed elasticity captures the tension increment clustering tendency which exists in real tension responses, while in stress-free elasticity this clustering is not captured.

\subsection{Shear modulus of prestressed amorphous solids}
The shear modulus $G$ can be obtained from the shear component of the virial stress.
All the contributions to the virial stress in Eq.~\eqref{EQ:Virial} can be obtained from the SSSs projection calculation (which directly gives $t^{\parallel}_b,t^{\perp}_b$ of all bonds $b$) via
\begin{align}
    \vec{t}_b&=(t_{b,p}+t^{\parallel}_b) \hat{R}^{\parallel}_{b,0}+ t^{\perp,1}_b \hat{R}^{\perp,1}_{b,0}+ t^{\perp,2}_b \hat{R}^{\perp,2}_{b,0},\\
    \vec{R}_b&= \left(
    R_{b,R} + \frac{|\vec{t}_b|}{k_b}
    \right) \hat{t}_b,
\end{align}
where 
%$\hat{b}^{\parallel}_0,\hat{b}^{\perp}_0$ are the unit vectors along and perpendicular to  bond $b$ in the reference state, and 
$\hat{t}_b\equiv \vec{t}_b/|\vec{t}_b|$ is the direction of the bond after the deformation. Comparing with the continuum limit we discussed at Eq.~\eqref{EQ:Ksigma}, this virial stress tensor is both a spatial average over the whole system, and uses the bond directions after deformation $\vec{R}_b$ (which is along the calculated $\vec{t}_b$) instead of original directions $\vec{R}_{b,0}$.  In the terminology of structural mechanics, this virial stress [Eq.~\eqref{EQ:Virial}] is the Cauchy stress (true stress, in the thermodynamic limit), and Eq.~\eqref{EQ:Ksigma} defines the Second Piola-Kirchhoff stress which uses bond directions in the undeformed state.  [In practice, using Eq.~\eqref{EQ:Ksigma} would just result in slightly higher errors.]

\begin{figure}[H]
    \centering
    \includegraphics[width=0.5\textwidth]{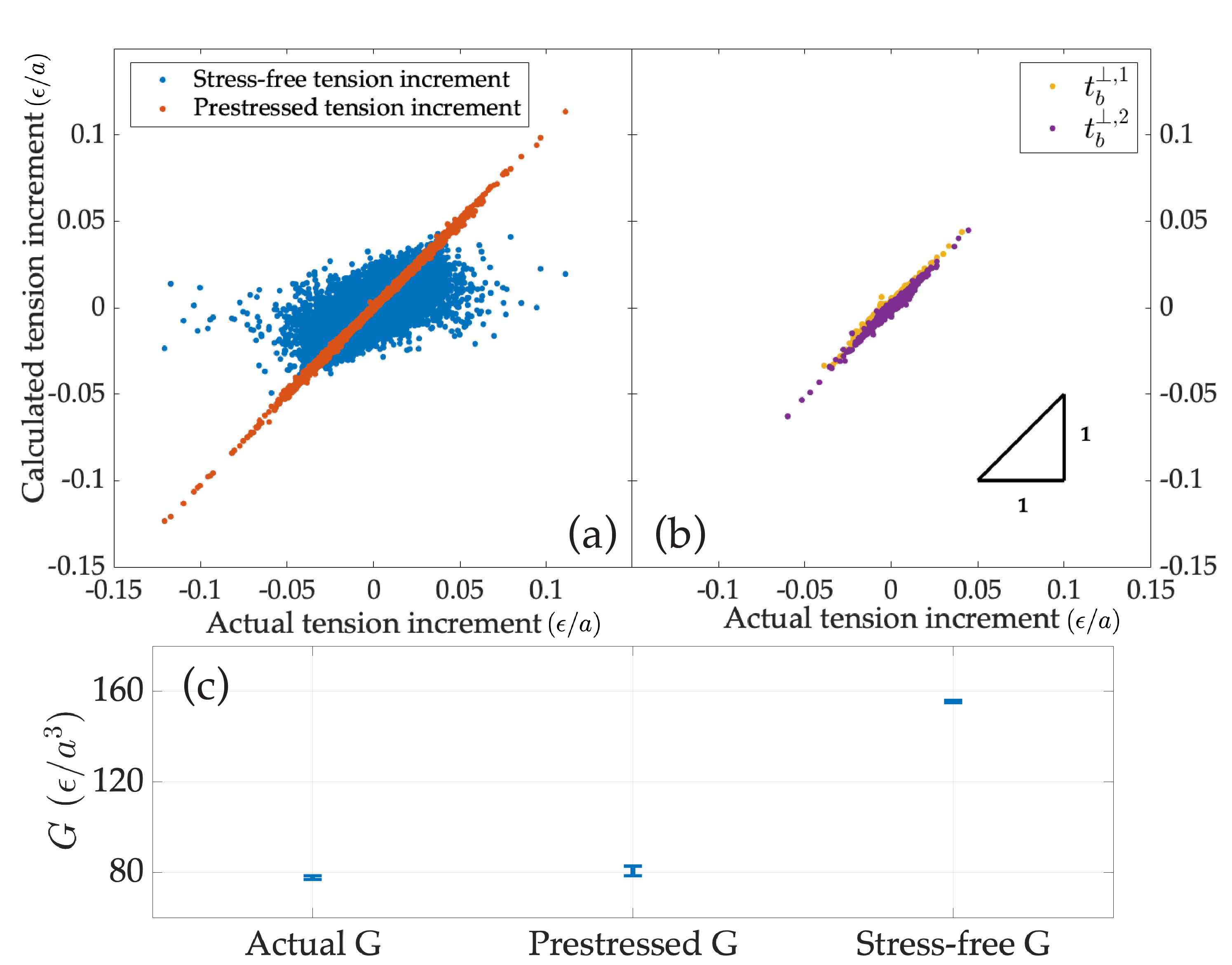}
    \caption{Comparing tension increment calculated from SSSs projection and numerical experiments (actual tension) of quasistatic $0.01\%$ shear strain [system size N = 10976 and cooling rate $\Gamma = 5\times 10^{-6}~\epsilon/(k_B\tau_0)$].  
    (a-b) Comparison of calculated and actual tension increment ($t^{\parallel},t^{\perp}$) of each bond, where good agreement is found in the case of prestressed formulation. 
    (c) Shear modulus $G$ from stress-free response and prestressed response [Eq.~\eqref{EQ:Virial}] compared to the actual $G$. The shear modulus is averaged over three different configurations at quasistatic $0.01\%$ shear strain, with the same system size and preparation protocol to the system in (a-b).
    }
    \label{Fig:tension_inc_predict}
\end{figure}

In Fig.~\ref{Fig:tension_inc_predict} we show the agreement between the actual and the calculated tension increment using the prestressed formulation on the bonds, as well as the agreement of the shear modulus. In contrast, the tension increment calculated from the stress-free formulation displays clear deviations.

This effect can also be viewed from the continuum elasticity formulation in Sec.~\ref{SEC:Continuum}: when the macroscopic shear deformation field $\partial _y u_x$ is plugged into the continuum theory, the elastic energy density is
\begin{equation}
    E/V = \frac{1}{2} (K_{xyxy}+\sigma_{p,yy} ) (\partial_y  u_x)^2 ,
\end{equation}
leading to a prestress-corrected shear modulus of $G=K_{xyxy}+\sigma_{p,yy} $. It is straightforward from this relation that an isotropic compressional prestress ($\sigma_{ij}=-p \delta_{ij}$) decreases the shear rigidity of a network.

The results shown so far demonstrates that including prestress effects is important to characterize the mechanical response of the model amorphous solids obtained from the simulations. We next use the configurations of the model amorphous solids to collect statistical information on the SSSs when different preparation protocols, i.e. different cooling rates, are utilized.

\subsection{General statistics of states of self-stress}\label{SEC:generalstatistics}
By varying the cooling rate as described above, we prepare particle assemblies that have different amount of disorder and compressive prestress, and have shown elsewhere that the higher cooling rates lead to a higher degree of disorder, characterized through a Voronoi analysis of the local particle packing which shows a decrease in the local icosahedral order \cite{vasisht2020emergence, vasisht2020computational}.

\begin{figure}%[H]
    \centering
    \includegraphics[width=0.5\textwidth]{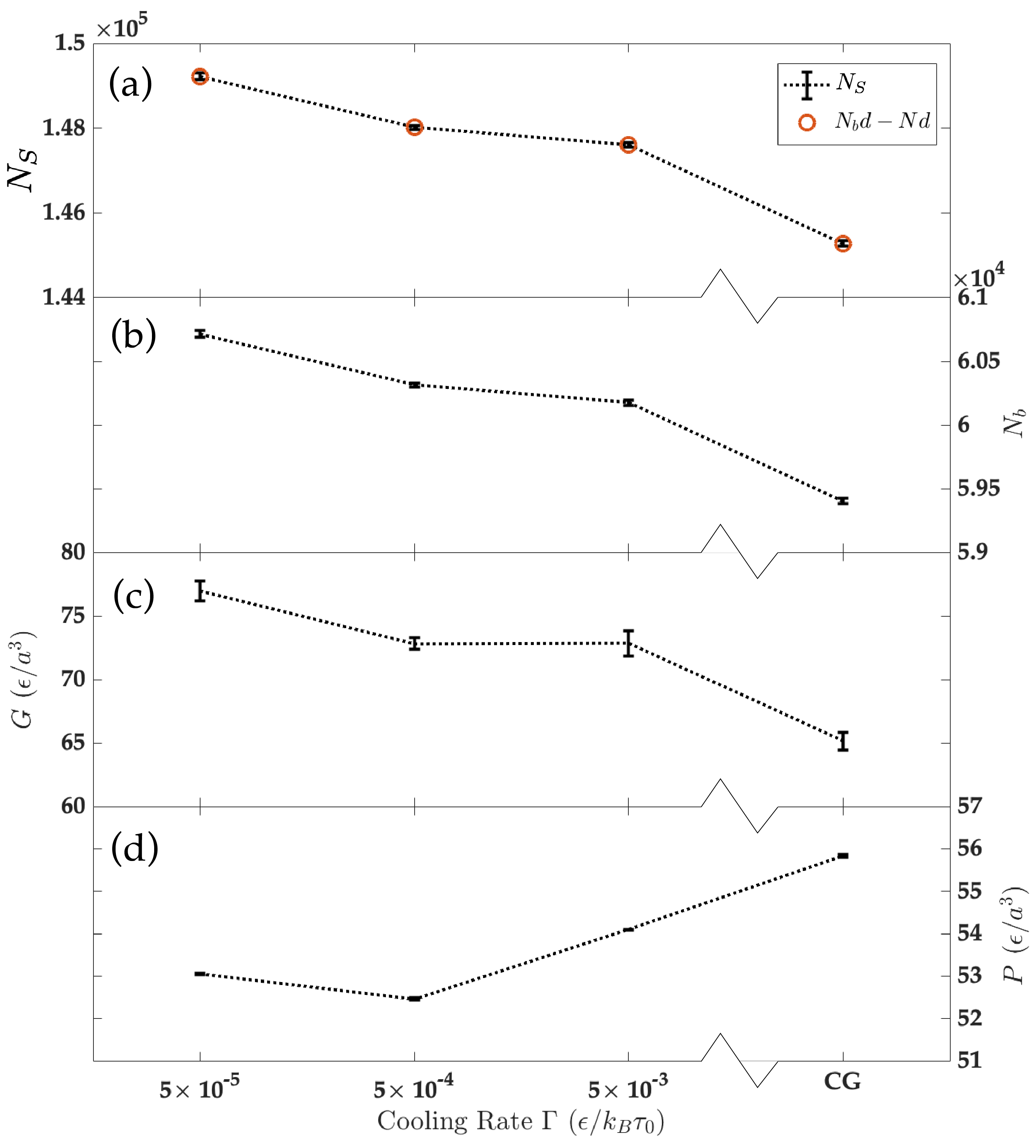}
    \caption{Statistics of SSSs at different cooling rate.  
    (a) Number of SSSs ($N_{\textrm{S}}$), (b) number of bonds ($N_b$), (c) measured shear modulus ($G$), (d) averaged normal stress ($P$) for different cooling rates $\Gamma$. Those quantities are averaged over 5 different configurations for each cooling rate, and the system size is $N = 10976$.)
    }\label{FIG:Numbers}
\end{figure}

The total number of SSSs in a prestressed system $N_S$ is directly related to the numbers of degrees of freedom and constraints via the Maxwell-Calladine index theorem [Eq.~\eqref{EQ:prestressedMC}] as the system exhibits no ZMs except for the trivial translations. In Fig.~\ref{FIG:Numbers}(a) we show the total number of SSSs in different configurations of our model amorphous solid that were prepared by varying the cooling rate. The number $N_{S}$ decreases with the increase of cooling rate. Figs.~\ref{FIG:Numbers}(b)-(d) show that this effect is accompanied by a decrease in the number of bonds and a decrease of the shear modulus, which is also associated to a net increase of the average compressional prestress. These findings support the idea that varying the preparation protocol by varying the cooling rate, as we do, induces different amount of prestress into amorphous solid configurations, which can be directly captured by the SSSs analysis. The data also indicate that more aged samples, which tend to be stiffer, feature larger amounts of SSSs. This is consistent with the finding in Ref.~\cite{vasisht2020emergence} that more aged samples (i.e., prepared with a lower cooling rate) contained larger amounts of (overconstrained) icosahedrally packed domains, identified through the Voronoi analysis of the particle packing. These domains, which tend to be stiffer, were shown to favor the accumulation of stress and promote dilation when the samples were driven towards yielding under a shear deformation. The comparison of the results obtained here with the analysis performed in Refs.~\cite{vasisht2020emergence, vasisht2020computational} support the idea that those phenomena, i.e., the increase in stiffness and in the tendency to accumulate stress and dilate under shear, should be due to prestress, and that sizable changes in the number of SSSs, as a result of sizable differences in prestress, eventually determine sizable changes in the linear, and even nonlinear, response of amorphous solids.

\subsection{Dipole stiffness in prestressed glasses}
SSSs have been also discussed in the context of the
mechanical response of amorphous solids to local perturbations (e.g., force dipoles), which could be connected to the localized or quasi-localized plastic processes that are key ingredients of the mechanical response of this class of materials~\cite{sussman2016spatial,lerner2018quasilocalized}. 
The analysis of nonaffine { and plastic} microscopic displacements under shear and of their spatial correlations has led to the notion of ``soft spot'', i.e., the presence of localized regions that are especially susceptible to plastic rearrangements under an external force\cite{Leonforte2004ContinuumI,Leonforte2004ContinuumII,maloney, chen2011measurement,rottler-prx2014}. 
%Studies of the mechanical properties of glassy materials often focus on the relationship between the structure and the dynamics under applied strains. In addition to approaches that attempt to find locally favored strutures, there are approaches that attempt to identify "soft spots", localized regions that are especially susceptible to plastic rearrangements due to external forces or thermal fluctuations. 

\begin{figure*}%[H]
    \centering
    \includegraphics[width=0.95\textwidth]{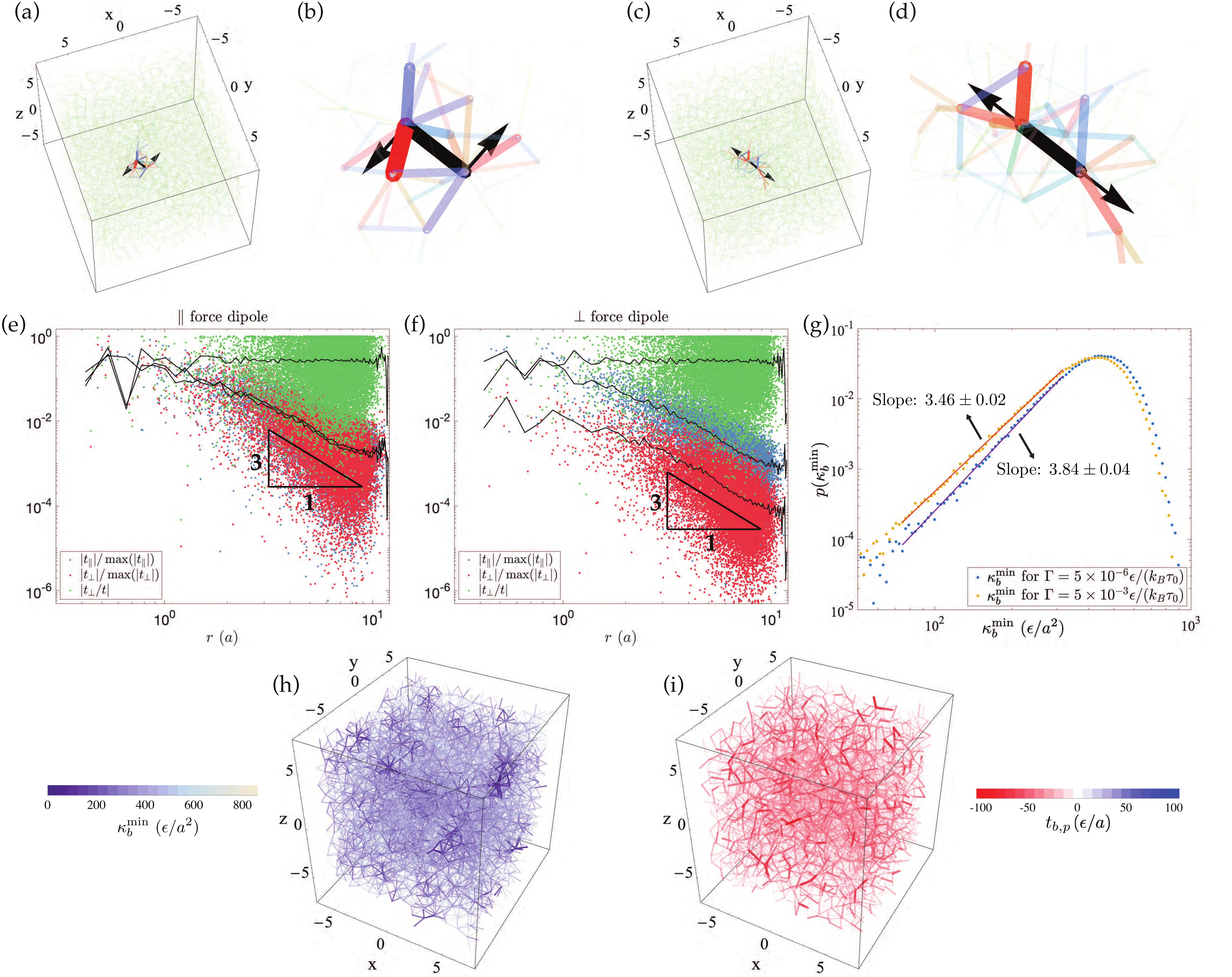}
    \caption{Dipole responses in amorphous solids. 
    (a-d) Tension responses to a transverse (a) or longitudinal (c) force dipole applied on the a bond (black arrows). The color scheme showing $t^{\parallel},t^{\perp}$ is the same as the one used in Fig.~\ref{FIG:sample-transverse-SSS}. (b) and (d) are zoomed-in views near the bond in (a) and (c) respectively.  
    (e-f) Spatial decay of tension responses for the two force dipoles in (a) and (c) respectively. 
   (g) Distribution of minimum dipole stiffness $\kappa_{b}^{\textrm{min}}$ for systems with different preparation cooling rates. System size is $N = 2916$. The distribution is averaged over 10  samples for each  cooling rate.
   (h) Spatial distribution of $\kappa_{b}^{\textrm{min}}$ on bonds for a configuration of system size $N=2916$ and preparation cooling rate $\Gamma = 5\times 10^{-6}~\epsilon/(k_B\tau_0)$. {Bonds with soft $\kappa_{b}^{\textrm{min}}$ are  thicker and more opaque.} (i) Pretension $t_{b,p}$ on bonds for the same configuration in (h).
   }
    \label{FIG:force_dipole}
\end{figure*}

The local response to force dipoles that we have discussed in Sec.~\ref{SEC:ProjectionDipole} has been directly related to these soft spots in amorphous solids \cite{sussman2016spatial,lerner2018quasilocalized,rainone2020pinching} and such response analysis has pointed to quasilocalized excitations (QLEs). In general, one needs to follow the time evolution of a system under an external deformation to identify the QLEs, however similar characteristics have been extracted, in the case of model glasses and amorphous solids, even from the low-frequency limit of the vibrational spectrum, i.e., from the linear response regime \cite{Leonforte2004ContinuumI,chen2011measurement,rottler-prx2014,Mosayeby2014soft}. 

Here we use the newly developed methods to examine the response of the model amorphous solids obtained in simulations to dipole forces, and collect the statistics of the local stiffness these systems display. In particular, we apply the approach for force dipoles discussed in Sec.~\ref{SEC:ProjectionDipole}, and calculate the stiffness as defined in Eq.~\eqref{EQ:DipoleK} for each bond, to identify regions which are soft. Measures of this kind lend themselves, in fact, to identify soft regions with respect to various external perturbations.

Interestingly, the relative displacement that a pair of
particles undergoes during a plastic rearrangement is typically not along the direction of the bond that initially connects them. Instead, this local deformation involves several sliding motions. Therefore, a particularly interesting measure of local softness, potentially complementing recently proposed ones \cite{lerner2018quasilocalized,Cubuk2015Identifying,Richard2020Predicting}, can be obtained by quantifying the stiffness against a dipole, \emph{whose forces are along a different direction from the one of the bond on which they are applied.} Thus, for each bond, in addition to computing the dipole stiffness to longitudinal (along $t^{\parallel}$) and transverse (along $t^{\perp}$) forces, 
we compute the softest dipole stiffness, $\kappa_b^{\textrm{min}}$, by minimizing it with respect to the direction of the dipole forces 
(details in App.~\ref{APP:Dipole}).

Fig.~\ref{FIG:force_dipole}(a) shows an example of applying longitudinal  and  transverse  force dipoles on a bond in our prestressed particle configurations, showing both $\parallel$ and $\perp$ tension responses of all bonds. 
%In particular, the stiffness to transverse force dipoles are a new measure we introduce here to probe local mechanical responses of amorphous solids. 
Remarkably, although the near field response is mainly in the $\parallel$ direction, the far field response contains significant $t^{\perp}$ components. 
%{\color{red} In fact, }
This is a genuine effect of prestress---these transverse stress response can only be captured when prestress is included.  
%, which are studied at the first time, indicate 
%the effect of prestress to determine the responses of soft solids. The near field of the local force dipole has larger magnitude of responses in $\parallel$ bond directions while in the far field, the prestress induces more bond rotations.

We collect the statistics of the minimum dipole stiffness $\kappa^{\textrm{min}}_b$ on bonds in the networks obtained with
%The minimized local dipole stiffness $\kappa_{min}$ is compared in 
two different preparation cooling rates as shown in Fig.~\ref{FIG:force_dipole}(g).   {These distributions feature a power-law tail for small $\kappa^{\textrm{min}}_b$, where $p(\kappa^{\textrm{min}}_b)\sim (\kappa^{\textrm{min}}_b)^{\beta}$ and $\beta$ appears to depend on the cooling rate.  
}
The results demonstrate the impact of preparation history to the mechanical stiffness in prestressed systems. 
{
The case with faster cooling rate leads to a smaller exponent and thus the distribution extends more to soft $\kappa^{\textrm{min}}_b$, agreeing with the general trend discussed in Sec.~\ref{SEC:generalstatistics}, where faster cooling rate leaves more heterogeneities regarding stress response in the sample.  Our results also provide an interesting comparison to (longitudinal) dipole stiffness distributions studied in Ref.~\cite{rainone2020statistical} for different models of computational glasses.
%the exponent we observe here exhibit a different exponent, due to the fact that small $\kappa^{\textrm{min}}_b$ values are dominated by transverse dipole responses, as well as very different preparation protocols and particle-particle interactions. 
}

{
Furthermore, we also characterize the spatial distribution of $\kappa^{\textrm{min}}_b$ and compare it to the spatial distribution of prestress $t_{b,p}$ (Fig.~\ref{FIG:force_dipole}h,i). Remarkably, $\kappa^{\textrm{min}}_b$ shows interesting spatial patterns whose possible correlations with the amount of prestress in these regions are however hard to tease out. Relating regions with clustered soft  $\kappa^{\textrm{min}}_b$  with soft spots will be an interesting question for future studies.  
}

%{\color{blue} We should comment more about the differences/similarities and shape of the functions?}
%From the statistics collected, $\bar{\kappa}_{min}(\Gamma = 5\times 10^{-6}) = 422.90\pm2.13 ~\epsilon/a^2; \bar{\kappa}_{min}(\Gamma = 5\times 10^{-3}) = 395.19\pm1.68 ~\epsilon/a^2$. An exponent is extracted in small $\kappa$ region as shown as fitted lines in Fig.~\ref{FIG:force_dipole}(c). Noting the power exponent as $b$, we have $b(\Gamma = 5\times 10^{-6}) = 3.84\pm0.04; b(\Gamma = 5\times 10^{-6}) = 3.46\pm0.02$.

\section{Discussion}\label{SEC:Conclusion}
Through this paper, we have developed a set of mathematical tools to investigate elasticity of prestressed discrete networks, which allows to disentangle the prestress from the structural disorder, 
%{\color{blue} this is not so clear for me, what about: "which allows to disentangle the prestress from the structural disorder},  
offering new insight into the mechanical response of amorphous solids.  

Our theory is based on the equilibrium and compatibility matrices, which rigorously defines SSSs and ZMs in mechanical networks. %This sentence is not clear for me: 
SSSs provide a linear space for the ensemble of force-balanced prestress distributions of the network, therefore capturing the possibility that %allowing 
the prestress may vary without altering the network configuration. 
Any mechanical network with SSSs can become prestressed, e.g., by introducing mismatching bond lengths (geometric frustration), which naturally arises 
%whereas prestress and SSSs naturally occurs 
in the solidification of amorphous solids in general. These 
We have generalized the equilibrium and compatibility matrices to these prestressed systems, which amounts to include additional constraints on bonds from prestress.  We have then used the new mathematical tools to analyze the mechanical response of prestressed triangular lattices and model amorphous solids obtained through 3D numerical simulations, and found good agreement between the numerically measured response and results obtained from our methods, whereas the time-complexity of the computation is greatly reduced.

In particular, 
%we should better explain this in the text: 
%{\color{blue} in triangular lattices} 
we not only characterize how prestress affects vibrational modes, both stabilizing soft modes and introducing a new class of ZMs (type $B$, which appear due to competing positive and negative terms in the energy), but also reveal its profound role on how the network carries additional stress. By analyzing prestressed SSSs of the network, and projecting external load to this linear space, our methods conveniently map how external stress transmits in prestressed lattices and bond networks obtained from model amorphous solids. We show a number of intriguing effects which are unique to prestressed networks, including the spatial heterogeneity in the stress-response to homogeneous loads, the power-law distribution of the local stiffness and the isotropic nature of the long-range stress response to local perturbations such as force dipoles.

These results provide new tools and insight to unravel the complexity of mechanical properties of amorphous solids.  They also open new paths for further exploration. For example, the new measure of minimum dipole stiffness we propose is potentially a new direct probe to analyze soft spots in amorphous solids. Relating this measure with existing characterizations of soft spots, as well as yielding of amorphous solids, will be interesting topics for future studies.

Moreover, the ability to separately vary the prestress and the microscopic bond configuration in our theory offers a new perspective to analyze the dynamics of amorphous solids under strain. In real amorphous solids, from granular matter to colloids, particle-particle interactions are not simple harmonic springs. Instead, they evolve under stress in complicated and often unpredictable ways---micro-fracturing, crushing, sintering. This evolution is further complicated by the role of the solvent, where contacts between particles can be lubricated or frictional. Therefore, \emph{stress distributions can significantly change with little change in the configuration}. Our methods provide a convenient set of tools to examine how the mechanical response of these amorphous materials may evolve as the system explores an ensemble of states where only internal stress is changed.  

Prestress records the preparation history in amorphous solids and has an important role in directing the mechanical response beyond the linear regime. 
The implications of prestress and preparation history, in fact, have been highlighted with respect to the development of flow inhomogeneities upon yielding \cite{vasisht2020emergence} and in connection with the fundamental physics mechanisms controlling brittle or ductile yielding phenomena \cite{Ozawa2018random,Singh2020Brittle,Barlow2020ductile}. Our methods, therefore, have wider relevance as they can be applied beyond the linear response to investigate, through the SSSs characteristics, the role of prestress in plasticity and yielding. The new tools discussed here can potentially close the feedback loop between structure and stress, opening a new pathway to study how amorphous solids yield and solidify under external strain.

Finally, friction plays a central role in the complex behaviors of exotic rigid states emerging in granular matter and dense suspensions, leading to fascinating phenomena from shear jamming to discontinuous shear thickening~\cite{Bi2011,seto2013discontinuous,wyart2014discontinuous,Lin2016Tunable,Guy2018Constraint,Liu2021,ramaswamy2021universal}. Interestingly, friction also lives in the $t^{\perp}$ channel of stress, and a linear relation with $e^{\perp}$ can also be assumed, hence it would be a natural next step to include friction into the SSSs description. However, the Coulomb threshold imposes an upper limit to $t^{\perp}$ that is not part of the methods developed here. How the frictional and the prestress contributions to $t^{\perp}$ interplay and affect the macroscopic dynamics of the material, will be of great interest for future studies.
%Furthermore, it would be a natural next step to include friction into the scenario of SSSs description of amorphous solids. 
%[add more discussions, granular, suspension, shear jamming]

\begin{acknowledgments}
This work is supported in part by the National Science Foundation under grant number NSF-DMR-2026825 (E.D.G and X.M.), NSF-EFRI-1741618 (S.Z., L.Z., and X.M.), 
and the Office of Naval Research under grant number MURI N00014-20-1-2479 (E.S. and X.M.)
\end{acknowledgments}

% {\color{red}*************Old notes*************}
% A few ideas for discussion
% \begin{itemize}
%     \item Efficiency of the computational method (discuss time complexity)
%     \item What is the main insight from this work: stress-bearing abilities of glasses, as characterized by  SSSs, provide  a new way to unravel mechanical response of glasses to any load, both macroscopic and local.  The statistics of these SSS will shed new light on understanding yielding, shear thinning and thickening, as well as providing input to field theory for the dynamics of dense suspensions.
%     \item Brittle-ductile crossover discussion?
% \end{itemize}

% % {\color{red}
% % type B ZMs

% % clarification of stress bearing and SSS

% % new measure of local stiffness--soft spots
% % }

% Mention future directions
% \begin{itemize}
%     \item How SSS evolves as the system is sheared (especially when the geometry is almost intact but prestress evolved)
%     \item Systems close to 
% \end{itemize}

% \vspace{5cm}

\appendix

%{\color{orange}
%\section{Global rotations in stressed elasticity}
%}

\section{SSSs formulation for shear response of prestressed networks}\label{APP:Projection}
\subsection{Projection of a shear load to the SSSs linear space}
When a mechanical network is subject to a shear load, the resulting bond extensions and rotations can be written as the sum of contributions from the affine shear field $e_{\textrm{affine}}$ and the nonaffine displacements $\mathbb{C}| u_{\text{rsp}} \rangle$,
\begin{align}
| e \rangle &= | e_{\textrm{affine}}  \rangle + 
\mathbb{C}| u_{\text{rsp}} \rangle,
\end{align}
which is similar to Eq.~\eqref{EQ:eparaANA} in the main text, but here we include both the parallel and the perpendicular components of $e$.  

When force-balance is reached, net force on each particle vanishes,
\begin{align}
- | f \rangle &=\mathbb{Q}|t\rangle 
= \mathbb{Q}\mathbb{K}(| e_{\text{affine}} \rangle + \mathbb{C} | u_{\text{rsp}} \rangle) = 0. 
\end{align}
This means that $|t\rangle = \mathbb{K}(| e_{\text{affine}} \rangle + \mathbb{C} | u_{\text{rsp}} \rangle)$ must be a vector that belongs to the null space of $\mathbb{Q}$.  Thus it can be written as  a linear combinations of the SSSs of the system.

To facilitate the discussion of this SSSs linear combination, we define the following notations.  
Let $\{\displaystyle \vec{t}_s^{~(1)} ,\ldots ,\vec{t}_s^{~({N_\textrm{S}})}\}$ be an orthonormal basis of the null space of $\mathbb{Q}$,
%the subspace ${\displaystyle \ker(\mathbb{Q})}$, 
and let ${\displaystyle P^Q_s}$ denote the $N_bd \times N_\textrm{S}$ matrix whose columns are $\displaystyle \vec{t}_s^{~(1)} ,\ldots ,\vec{t}_s^{~({N_\textrm{S}})}$, i.e., ${\displaystyle P^Q_s={\begin{bmatrix}\vec{t}_s^{~(1)} ,\ldots ,\vec{t}_s^{~({N_\textrm{S}})}\end{bmatrix}}}$. 
One can also define the $N_b d \times (N_b d - N_\textrm{S})$ dimensional matrix ${\displaystyle P^Q_r}$ whose columns are an orthonormal basis of the orthogonal compliment of the null space of $\mathbb{Q}$.  
%subspace $\textrm{range}{(\mathbb{Q})}$; 
Similarly one can define 
the $Nd \times N_\textrm{0}$ matrix ${\displaystyle P^C_s}$ whose columns are an orthonormal basis of the null space of $\mathbb{C}$ (ZMs), 
%subspace $\ker{(\mathbb{C})}$; 
and the $Nd \times (Nd-N_\textrm{0})$ matrix ${\displaystyle P^C_r}$ whose columns are an orthonormal basis of the orthogonal compliment of the null space of $\mathbb{C}$. 
%subspace $\textrm{range}{(\mathbb{C})}$. 
These matrices are represented as,
\begin{align}
P^Q_s &= 
{\begin{bmatrix}\vec{t}_s^{~(1)} ,\ldots ,\vec{t}_s^{~({N_\textrm{S}})}\end{bmatrix}},\\
P^Q_r &= 
{\begin{bmatrix}\vec{t}_r^{~(1)} ,\ldots ,\vec{t}_r^{~({N_b d - N_\textrm{S}})}\end{bmatrix}},\\
P^C_s &= 
{\begin{bmatrix}\vec{u}_s^{~(1)} ,\ldots ,\vec{u}_s^{~({N_\textrm{0}})}\end{bmatrix}},\\
P^C_r &= 
{\begin{bmatrix}\vec{u}_r^{~(1)} ,\ldots ,\vec{u}_r^{~({Nd -N_\textrm{0}})}\end{bmatrix}},
\end{align}
and they satisfy the following identities,
\begin{align}
(P^Q_s)^T \cdot P^Q_s &= \mathbb{I}_{(N_\textrm{S})} ,\\
(P^Q_r)^T \cdot P^Q_r &= \mathbb{I}_{(N_b d - N_\textrm{S})}, \\
(P^C_s)^T \cdot P^C_s &= \mathbb{I}_{(N_{0})}, \\
(P^C_r)^T \cdot P^C_r &= \mathbb{I}_{(Nd-N_{0})}, \\
P^Q_s \cdot (P^Q_s)^T + P^Q_r \cdot (P^Q_r)^T &= \mathbb{I}_{(N_b d)},\label{EQ:INb} \\
P^C_s \cdot (P^C_s)^T + P^C_r \cdot (P^C_r)^T &= \mathbb{I}_{(N d)},
\end{align}
where $\mathbb{I}_{(N_\textrm{S})}$ is an identity matrix of dimension $N_\textrm{S}$, and other definitions follows similarly.

As we discussed above, $|t\rangle$ is a linear combination of the SSSs, 
\begin{align}
| t \rangle  &= \sum_{i}^{N_\textrm{S}} \alpha_i | t_\textrm{SSS, i} \rangle = P_s^{Q} \cdot \vec{\alpha},
\end{align}
where $\vec{\alpha}$ are coefficients of the linear combination of $|t\rangle$ as the SSSs.

Because the basis we use are orthonormal, 
\begin{align}
\vec{\alpha} &= (P^{Q}_s)^T | t \rangle = (P^{Q}_s)^T \mathbb{K} (| e_{\text{affine}} \rangle + \mathbb{C} | u_{\text{rsp}} \rangle), \label{EQ:tension_proj_alpha}\\
\vec{0} &= (P^{Q}_r)^T | t \rangle = (P^{Q}_r)^T \mathbb{K} (| e_{\text{affine}} \rangle + \mathbb{C} | u_{\text{rsp}} \rangle), \label{EQ:tension_force_equil}
\end{align}
inserting identity matrix [Eq.~\eqref{EQ:INb}] into Eq.~\eqref{EQ:tension_force_equil} and using the fact that,
\begin{align*}
\mathbb{Q} \cdot P_s^Q = \vec{0},
\end{align*}
we have,

\begin{widetext}

\begin{align}
\vec{0} &= (P^Q_r)^T \mathbb{K}\cdot [P^Q_s \cdot (P^Q_s)^T + P^Q_r \cdot (P^Q_r)^T] \cdot (| e_{\text{affine}} \rangle +  \mathbb{C} | u_{\text{rsp}} \rangle) \notag\\
&= (P^Q_r)^T \mathbb{K}\cdot [P^Q_s \cdot (P^Q_s)^T \cdot \mathbb{Q}^T + P^Q_r \cdot (P^Q_r)^T \cdot \mathbb{Q}^T] \cdot | u_{\text{rsp}} \rangle \notag\\ 
& ~~~ + (P^Q_r)^T \mathbb{K}\cdot [P^Q_s \cdot (P^Q_s)^T + P^Q_r \cdot (P^Q_r)^T] \cdot | e_{\text{affine}} \rangle \notag\\
&= (P^Q_r)^T \mathbb{K}\cdot P^Q_r \cdot (P^Q_r)^T \cdot \mathbb{Q}^T \cdot | u_{\text{rsp}} \rangle + (P^Q_r)^T \mathbb{K}\cdot [P^Q_s \cdot (P^Q_s)^T + P^Q_r \cdot (P^Q_r)^T] \cdot | e_{\text{affine}} \rangle \notag\\
&= \mathbb{K}_{rr} (P^Q_r)^T \mathbb{Q}^T \cdot | u_{\text{rsp}} \rangle + [\mathbb{K}_{rs} (P^Q_s)^T + \mathbb{K}_{rr} (P^Q_r)^T] \cdot | e_{\text{affine}} \rangle \notag\\
& \implies
(P^Q_r)^T \mathbb{Q}^T \cdot | u_{\text{rsp}} \rangle = - (\mathbb{K}_{rr})^{-1} [\mathbb{K}_{rs} (P^Q_s)^T + \mathbb{K}_{rr} (P^Q_r)^T] \cdot | e_{\text{affine}} \rangle,
\label{zero}
\end{align}
where we defined the decomposition of $\mathbb{K}$ into the null and orthogonal compliment space as
\begin{align}
\mathbb{K} \rightarrow
\begin{pmatrix}
(P^Q_s)^T\cdot\mathbb{K}\cdot P^Q_s & (P^Q_s)^T\cdot\mathbb{K}\cdot P^Q_r\\
(P^Q_r)^T\cdot\mathbb{K}\cdot P^Q_s & (P^Q_r)^T\cdot\mathbb{K}\cdot P^Q_r\\
\end{pmatrix}
=
\begin{pmatrix}
\mathbb{K}_{ss} & \mathbb{K}_{sr}\\
\mathbb{K}_{rs} & \mathbb{K}_{rr}\\
\end{pmatrix}
\end{align}
and also used  the fact that $\mathbb{K}_{rr}$ is invertible.

The coefficients $\vec{\alpha}$ in Eq.~\eqref{EQ:tension_proj_alpha} can then be solved as
\begin{align*}
\vec{\alpha} &= (P^Q_s)^T \mathbb{K}\cdot [P^Q_s \cdot (P^Q_s)^T + P^Q_r \cdot (P^Q_r)^T] \cdot (\mathbb{Q}^T | u_{\text{rsp}} \rangle +  | e_{\text{affine}} \rangle)\\
&= (P^Q_s)^T \mathbb{K}\cdot P^Q_r \cdot (P^Q_r)^T \cdot \mathbb{Q}^T \cdot | u_{\text{rsp}} \rangle\\
&~~~ + (P^Q_s)^T \mathbb{K}\cdot [P^Q_s \cdot (P^Q_s)^T + P^Q_r \cdot (P^Q_r)^T] \cdot | e_{\text{affine}} \rangle\\
&= \mathbb{K}_{sr} (P^Q_r)^T \mathbb{Q}^T \cdot | u_{\text{rsp}} \rangle + [\mathbb{K}_{ss} (P^Q_s)^T + \mathbb{K}_{sr} (P^Q_r)^T] \cdot | e_{\text{affine}} \rangle.
\end{align*}
Plug in Equation~\eqref{zero} to eliminate $u_{\text{rsp}}$,
\begin{align}
\vec{\alpha}
&= \mathbb{K}_{sr} \{ - (\mathbb{K}_{rr})^{-1} [\mathbb{K}_{rs} (P^Q_s)^T + \mathbb{K}_{rr} (P^Q_r)^T] \cdot | e_{\text{affine}} \rangle \}
+ [\mathbb{K}_{ss} (P^Q_s)^T + \mathbb{K}_{sr} (P^Q_r)^T] \cdot | e_{\text{affine}} \rangle \notag\\
&= \{ \mathbb{K}_{sr} (P^Q_r)^T + \mathbb{K}_{ss} (P^Q_s)^T - \mathbb{K}_{sr} (\mathbb{K}_{rr})^{-1} [\mathbb{K}_{rs} (P^Q_s)^T + \mathbb{K}_{rr} (P^Q_r)^T] \} \cdot | e_{\text{affine}} \rangle \notag\\
&= [ \mathbb{K}_{ss} (P^Q_s)^T - \mathbb{K}_{sr} (\mathbb{K}_{rr})^{-1} \mathbb{K}_{rs} (P^Q_s)^T ] \cdot | e_{\text{affine}} \rangle \notag\\
&= [ \mathbb{K}_{ss} - \mathbb{K}_{sr} (\mathbb{K}_{rr})^{-1} \mathbb{K}_{rs} ] \cdot (P^Q_s)^T \cdot | e_{\text{affine}} \rangle .
\end{align}

\end{widetext}

This can be further simplified by 
letting $\mathbb{A} = \mathbb{K}^{-1}$, and decompose $\mathbb{A}$ into the column-space and null-space  of $\mathbb{Q}$ as,
\begin{align}
\mathbb{A} \rightarrow
\begin{pmatrix}
(P^Q_s)^T\cdot\mathbb{A}\cdot P^Q_s & (P^Q_s)^T\cdot\mathbb{A}\cdot P^Q_r\\
(P^Q_r)^T\cdot\mathbb{A}\cdot P^Q_s & (P^Q_r)^T\cdot\mathbb{A}\cdot P^Q_r\\
\end{pmatrix}
=
\begin{pmatrix}
\mathbb{A}_{ss} & \mathbb{A}_{sr}\\
\mathbb{A}_{rs} & \mathbb{A}_{rr}\\
\end{pmatrix}.
\end{align}
One can see that,
\begin{align}
\mathbb{K}_{ss} \cdot \mathbb{A}_{ss} + \mathbb{K}_{sr} \cdot \mathbb{A}_{rs} &= \mathbb{I}_{ss} ,\\
\mathbb{K}_{rs} \cdot \mathbb{A}_{ss} + \mathbb{K}_{rr} \cdot \mathbb{A}_{rs} &= \vec{0}_{rs} .
\end{align}
Right multiply by $(\mathbb{A}_{ss})^{-1}$ on both sides of two equations,
\begin{align}
\mathbb{K}_{ss} + \mathbb{K}_{sr} \cdot \mathbb{A}_{rs} \cdot (\mathbb{A}_{ss})^{-1} = (\mathbb{A}_{ss})^{-1} ,\\
\mathbb{K}_{rs} = -\mathbb{K}_{rr} \cdot \mathbb{A}_{rs} \cdot (\mathbb{A}_{ss})^{-1} .
\end{align}
Then,
\begin{align}
\mathbb{K}_{ss} + \mathbb{K}_{sr} \cdot \mathbb{A}_{rs} \cdot (\mathbb{A}_{ss})^{-1} = (\mathbb{A}_{ss})^{-1} , \\
-(\mathbb{K}_{rr})^{-1} \cdot \mathbb{K}_{rs} = \mathbb{A}_{rs} \cdot (\mathbb{A}_{ss})^{-1} .
\end{align}
Combining these two equations we have,
\begin{align}
((\mathbb{K}^{-1})_{ss})^{-1} &= (\mathbb{A}_{ss})^{-1}\\
&= \mathbb{K}_{ss} - \mathbb{K}_{sr} \cdot (\mathbb{K}_{rr})^{-1} \cdot \mathbb{K}_{rs}.
\end{align}

As a result, $\vec{\alpha}$ is simplified to
\begin{align}
\vec{\alpha}
= ((\mathbb{K}^{-1})_{ss})^{-1} \cdot (P^Q_s)^T \cdot | e_{\text{affine}} \rangle
\label{EQ:alpha-coeff}
\end{align}
and the tension response to this external shear is
\begin{align}
\label{tension-response}
  | t \rangle = P^Q_s \cdot ((\mathbb{K}^{-1})_{ss})^{-1} \cdot (P^Q_s)^T \cdot | e_{\text{affine}} \rangle .
\end{align}
Note that this $t$ includes both $t^{\parallel}$ and $t^{\perp}$, and this formulation applies to other types of homogeneous strain, such as hydrostatic compression, as well.

\subsection{Affine bond deformation in prestressed systems}
In this subsection we derive the $e_{\textrm{affine}}$ field for any external load represented by a strain tensor $\epsilon$.

We start from the (affine) deformation gradient 
\begin{align}
    \Lambda_{ij}\equiv \frac{\partial R_i}{\partial R_{0,j}}
\end{align}
where the strain tensor $\epsilon=(\Lambda^{T}\Lambda-I)/2$.  The affinely deformed positions of each particle are then 
\begin{align}
    \vec{R}_{\ell,\textrm{affine}} = \Lambda \cdot \vec{R}_{\ell,0}.
\end{align}
We can then use the formulation discussed in Sec.~\ref{SEC:Prestress} [Eq.\eqref{EQ:e2}] to calculate the affine extensions/rotations $e_{b,\text{affine}}^{\parallel}, e_{b,\text{affine}}^{\perp} $ for each bond, which consist $| e_{\text{affine}} \rangle$.

\section{Dipole stiffness $\kappa$ in prestressed systems}\label{APP:Dipole}
When a pair of dipole forces is applied on a mechanical network between two particles that belong to the same rigid cluster, the network will show a linear response with tension distributed on the bonds.  In this Appendix we derive  the stress field of a prestressed network in response to the force dipole, and obtain a computationally efficient formula that gives the stiffness the system has against this force dipole.

\subsection{Local dipole stiffness}
As discussed in Sec.~\ref{SEC:ProjectionDipole}, the sum of the external force dipole and the tension response, $| t_\textrm{dipole} \rangle + | t_\textrm{rsp} \rangle$, must be a SSS of the prestressed network, 
%This sum can always be expanded as a linear combination of all SSSs of the prestressed network,
\begin{align*}
     | t_\textrm{dipole} \rangle + | t_\textrm{rsp} \rangle 
     =\sum_{i}^{N_\textrm{SSS}} \alpha_i | t_\textrm{SSS, i} \rangle.
\end{align*}
The coefficients $\alpha_i$ are determined in the same way as discussed in App.~\ref{APP:Projection}, just by replacing $|e_{\textrm{aff}}\rangle$ with $|b\rangle$.  As a result
\begin{align*}
     | t_\textrm{dipole} \rangle + | t_\textrm{rsp} \rangle 
     =P^Q_s \cdot ((\mathbb{K}^{-1})_{ss})^{-1} \cdot (P^Q_s)^T \cdot | b \rangle .
\end{align*}

We can thus use this in the expression for the dipole stiffness [Eq.~\eqref{EQ:DipoleK}], where the denominator is now
\begin{align}
\langle{f_{\textrm{dipole}}|u_{\textrm{rsp}} }\rangle
&= - \langle b | \mathbb{K} \mathbb{C} | u_{\textrm{rsp}} \rangle 
=-\langle b | t_\textrm{rsp} \rangle 
\nonumber\\
    &= k_b - \displaystyle \sum_{i,j}^{N_{\textrm{SSS}}} \langle b |t_{\textrm{SSS},i}\rangle
    \lbrack (\mathbb{K}^{-1})_{ss} \rbrack^{-1}
    \langle t_{\textrm{SSS},j} | b \rangle,
\end{align}
where we used Eq.~\eqref{EQ:FDipole} and 
the fact that $| t_\textrm{rsp} \rangle =\mathbb{K} \mathbb{C} | u_{\textrm{rsp}} \rangle $ (bond extension causes tension).

Therefore the dipole stiffness is
\begin{align}
    \kappa_{b}
    &= \frac{ \langle{f_{\textrm{dipole}}|f_{\textrm{dipole}}}\rangle}{\langle{f_{\textrm{dipole}}|u_{\textrm{rsp}} }\rangle} \notag \\
    &= \frac{ \langle b | \mathbb{K} \mathbb{C} \mathbb{Q}\mathbb{K}| b \rangle }{ \langle b|\mathbb{K}|b\rangle - \langle b | \displaystyle \sum_{i,j}^{N_{\textrm{SSS}}} |t_{\textrm{SSS},i}\rangle
    \lbrack (\mathbb{K}^{-1})_{ss} \rbrack^{-1}
    \langle t_{\textrm{SSS},j} | b \rangle}
    \label{EQ:kappa}
    \\
    &= \frac{2k_b^2}{ k_b -  \displaystyle \sum_{i,j}^{N_{\textrm{SSS}}} \langle b |t_{\textrm{SSS},i}\rangle
    \lbrack (\mathbb{K}^{-1})_{ss} \rbrack^{-1}
    \langle t_{\textrm{SSS},j} | b \rangle} \notag ,
\end{align}
where the factor of 2 in the last line comes from the fact that $f_{\textrm{dipole}}$ involves 2 particles.

\subsection{Non-local dipole stiffness}

Besides applying the force dipole on an arbitrary existing bond $b$ in the system, one could also apply a force dipole between two particles which are not  connected.  As we discuss below, this pair of dipole introduces a new constraint into the system.  

Force balance with imposed force dipole $| f_\textrm{dipole} \rangle$ can be written as:
\begin{align}
    0 &= | f_\textrm{dipole} \rangle
    +| f_\textrm{rsp} \rangle = 
    | f_\textrm{dipole} \rangle - \mathbb{Q}\mathbb{K}\mathbb{C} | u_{\textrm{rsp}} \rangle, 
    %\\&= - \mathbb{Q} ( | t_\textrm{dipole} \rangle + \mathbb{K}\mathbb{C} | u \rangle )
\end{align}
where $| u_{\textrm{rsp}} \rangle$ indicates the particle displacements in response to the force dipole as the system reaches force balance. Unlike the local dipole case discussed above, here $f_\textrm{dipole}$ can not be written as $\mathbb{Q}| t_\textrm{dipole} \rangle $ on an existing bond, as the two sites are not connected.  Instead, we can introduce an auxiliary bond  between the two sites that carry $| f_\textrm{dipole} \rangle$. In this sense, a new constraint and thus a new SSS is added to the system by the auxiliary bond.

Here we introduce the new $\mathbb{Q},\mathbb{C}$ matrices after introducing the auxiliary bond (which has zero spring constant so that it will not induce tension responses) as
\begin{align*}
    &\tilde{\mathbb{C}} =
    \begin{pmatrix}
     \mathbb{C}\\
     C_a
    \end{pmatrix},
    &\tilde{\mathbb{Q}} =
    \begin{pmatrix}
     \mathbb{Q} & Q_a
    \end{pmatrix},
    &&\tilde{\mathbb{K}} =
    \begin{pmatrix}
     \mathbb{K} & \\
     & 0
    \end{pmatrix},\\
    &\tilde{\mathbb{C}} \cdot | u \rangle =
    \begin{pmatrix}
     | e \rangle\\
     e_a
    \end{pmatrix},
    &\tilde{\mathbb{Q}} \cdot
    \begin{pmatrix}
     | t \rangle\\
     t_a
    \end{pmatrix} = - | f \rangle.
\end{align*}
The dimension of the bond space is extended with one additional component from the auxiliary bond indexed as $a$.

Now the force-balanced total tension can be written as
\begin{align}
    | \tilde{t} \rangle = 
    \begin{pmatrix}
     | t_\textrm{rsp} \rangle\\
     t_\textrm{dipole}
    \end{pmatrix} =
    \begin{pmatrix}
     \mathbb{K}\mathbb{C}|u_\textrm{rsp}\rangle\\
     t_\textrm{dipole}
    \end{pmatrix} =
    \tilde{\mathbb{K}}\tilde{\mathbb{C}} | u_\textrm{rsp} \rangle + | \tilde{t}_\textrm{dipole} \rangle, 
\end{align}
where $t_\textrm{dipole}$ lives on the auxillary bond.  This $| \tilde{t} \rangle$ satisfies force balance and must be a SSS of the network with the auxillary bond.

Similar to App.~\ref{APP:Projection}, one can define $P_s^{\tilde{Q}}, P_r^{\tilde{Q}}, P_s^{\tilde{C}}, P_r^{\tilde{Q}}$ to the new  matrices. We can then decompose the total tension onto SSSs, 
\begin{align*}
| \tilde{t} \rangle  &= \sum_{i}^{N_\textrm{SSS}} \alpha_i | t_\textrm{SSS, i} \rangle = P_s^{\tilde{Q}} \cdot \vec{\alpha},
\end{align*}
where $\vec{\alpha}$'s are coefficients of the linear combination of the SSSs.

We can compute the tension in a similar way,
\begin{align*}
(P^{\tilde{Q}}_s)^T | \tilde{t} \rangle & = \vec{\alpha},\\
(P^{\tilde{Q}}_r)^T | \tilde{t} \rangle & = \vec{0}.
\end{align*}
Thus
\begin{align}
\vec{0} &= (P^{\tilde{Q}}_r)^T  \left( \tilde{\mathbb{K}}\tilde{\mathbb{C}} | u_\textrm{rsp} \rangle + | \tilde{t}_\textrm{dipole} \rangle \right)\\
&= (P^{\tilde{Q}}_r)^T \tilde{\mathbb{K}} \left( P^{\tilde{Q}}_r(P^{\tilde{Q}}_r)^T + P^{\tilde{Q}}_s(P^{\tilde{Q}}_s)^T \right) \tilde{\mathbb{Q}}^T | u_\textrm{rsp} \rangle 
\nonumber\\
&\quad + (P^{\tilde{Q}}_r)^T | \tilde{t}_\textrm{dipole} \rangle\\
&= (P^{\tilde{Q}}_r)^T \tilde{\mathbb{K}} P^{\tilde{Q}}_r(P^{\tilde{Q}}_r)^T \tilde{\mathbb{Q}}^T | u_\textrm{rsp} \rangle + (P^{\tilde{Q}}_r)^T | \tilde{t}_\textrm{dipole} \rangle\\
&= \tilde{\mathbb{K}}_{rr} (P^{\tilde{Q}}_r)^T \tilde{\mathbb{Q}}^T | u_\textrm{rsp} \rangle + (P^{\tilde{Q}}_r)^T | \tilde{t}_\textrm{dipole} \rangle.
\label{EQ:urspnld}
\end{align}
As a result, 
\begin{equation}
(P^{\tilde{Q}}_r)^T \tilde{\mathbb{Q}}^T | u_\textrm{rsp} \rangle =
- \left( \tilde{\mathbb{K}}_{rr} \right)^{-1} (P^{\tilde{Q}}_r)^T | \tilde{t}_\textrm{dipole} \rangle .
\label{EQ:proj_0}
\end{equation}
Note that although the spring constant of the auxiliary bond is 0, the matrix $\tilde{\mathbb{K}}_{rr}$ is still invertable.  This is because the auxiliary bond is a redundant bond (otherwise the network would yield, resulting in no stress in linear response), which increases the dimension of the SSSs space and does not introduce new vectors in the orthogonal compliment space. As a result, $\tilde{\mathbb{K}}_{rr}=\mathbb{K}_{rr}$ and is invertable. 

This can be used to find the coefficients for the SSSs
\begin{align}
    \vec{\alpha} &= (P^{\tilde{Q}}_s)^T | \tilde{t} \rangle
    \nonumber\\
    &= (P^{\tilde{Q}}_s)^T  \left( \tilde{\mathbb{K}}\tilde{\mathbb{C}} | u_\textrm{rsp} \rangle + | \tilde{t}_\textrm{dipole} \rangle \right)
    \nonumber\\
    &= \tilde{\mathbb{K}}_{sr} (P^{\tilde{Q}}_r)^T \tilde{\mathbb{Q}}^T | u_\textrm{rsp} \rangle + (P^{\tilde{Q}}_s)^T | \tilde{t}_\textrm{dipole} \rangle
    \nonumber\\
    &= \left( (P^{\tilde{Q}}_s)^T - \tilde{\mathbb{K}}_{sr} \left( \tilde{\mathbb{K}}_{rr} \right)^{-1} (P^{\tilde{Q}}_r)^T 
    \right) | \tilde{t}_\textrm{dipole} \rangle ,
\end{align}
where in the last line we used Eq.~\eqref{EQ:proj_0}. These coefficients gives the tension field in response to a nonlocal dipole.  

We can then proceed to calculate the dipole stiffness.  
Applying $P^{\tilde{Q}}_r$ on both sides of Eq.~\eqref{EQ:proj_0}, we have the left hand side, 
\begin{align}
P^{\tilde{Q}}_r (P^{\tilde{Q}}_r)^T \tilde{\mathbb{Q}}^T | u \rangle %&= - P^{\tilde{Q}}_r
%\left( \tilde{\mathbb{K}}_{rr} \right)^{-1} (P^{\tilde{Q}}_r)^T | \tilde{t}_\textrm{dipole} \rangle\\
%(\textrm{LHS}) 
&= \left( \mathbb{I} - P^{\tilde{Q}}_s (P^{\tilde{Q}}_s)^T \right) \tilde{\mathbb{Q}}^T | u \rangle\\
&= \tilde{\mathbb{Q}}^T | u \rangle - P^{\tilde{Q}}_s \left( \tilde{\mathbb{Q}} \cdot P^{\tilde{Q}}_s  \right)^T | u \rangle\\
&= \tilde{\mathbb{Q}}^T | u \rangle.
\end{align}
Equaling it to the right hand side we have
\begin{align}
    \tilde{\mathbb{Q}}^T | u \rangle 
    = - P^{\tilde{Q}}_r \left( \tilde{\mathbb{K}}_{rr} \right)^{-1} (P^{\tilde{Q}}_r)^T | \tilde{t}_\textrm{dipole} \rangle ,
\end{align}
expressing the displacement field as a function of the imposed dipole. 

One can then compute the non-local dipole stiffness $\kappa_a$ as (the force dipole  $|f_{\textrm{dipole}} \rangle = -\mathbb{\tilde{Q}} | \tilde{t}_\textrm{dipole} \rangle \equiv -\mathbb{\tilde{Q}} | a \rangle$ where 
$| a \rangle$ is the vector in the labeling space of bonds which has zeros in all bonds and unity on the $a$-th component (the auxiliary bond), and a spring constant of unity is added (which is only  used to represent the external force dipole, where the actual spring constant of the auxillary bond regarding its contribution to the network response is still 0 as discussed above),  
\begin{align}
    \kappa_{a}
    &=
    \frac{ \langle{f_{\textrm{dipole}}|f_{\textrm{dipole}}}\rangle}{\langle{f_{\textrm{dipole}}|u }\rangle}\\
    &= \frac{ \langle{\tilde{t}_{\textrm{dipole}}| 
    \mathbb{\tilde{Q}}^T \mathbb{\tilde{Q}}
    |\tilde{t}_{\textrm{dipole}}}\rangle}{-\langle{ \tilde{t}_\textrm{dipole} | \mathbb{\tilde{Q}}^T| u }\rangle}\\
    &=
    \frac{ \langle{\tilde{t}_{\textrm{dipole}}| 
    \mathbb{\tilde{Q}}^T \mathbb{\tilde{Q}}
    |\tilde{t}_{\textrm{dipole}}}\rangle}{\langle{ \tilde{t}_\textrm{dipole} | 
    P^{\tilde{Q}}_r \left( \tilde{\mathbb{K}}_{rr} \right)^{-1} (P^{\tilde{Q}}_r)^T | \tilde{t}_\textrm{dipole}
    }\rangle}\\
    &=
    \frac{ \langle{a | 
    \mathbb{\tilde{Q}}^T \mathbb{\tilde{Q}}
    |a }\rangle}{\langle{ a | 
    P^{\tilde{Q}}_r \left( \tilde{\mathbb{K}}_{rr} \right)^{-1} (P^{\tilde{Q}}_r)^T | a
    }\rangle}\\
    &=
    \frac{2}{\langle{ a | 
    P^{\tilde{Q}}_r \left( \tilde{\mathbb{K}}_{rr} \right)^{-1} (P^{\tilde{Q}}_r)^T | a
    }\rangle}
\end{align}

The local force dipole response is a special case of non-local force dipole response. When considering local force dipoles, the auxiliary bond $a$ overlaps with bond $b$. One can show that in this case, the non-local force dipole stiffness reduces to local force dipole stiffness.

\begin{widetext}

\subsection{Minimum local dipole stiffness $\kappa_{b,\textrm{min}}$ in prestressed systems}
In this section we derive the minimum dipole stiffness on any given bond $b$, which is the lowest with respect to the 
%To minimize $\kappa$ for a local force dipole on bond $b$ with respect to arbitrary 
combination of $\parallel$ and $\perp$ directions.  To do this we use $(\theta,\phi)$ to denote the direction of the dipole forces, and the resulting $t_{\textrm{dipole}}$ can be written as
%, one is to find
\begin{align}
    | t_{\text{dipole}} \rangle &=
    \begin{pmatrix}
     0\\
     \vdots\\
     \cos{\theta}\\
     \sin{\theta}\cos{\phi}\\
     \sin{\theta}\sin{\phi}\\
     \vdots\\
     0\\
    \end{pmatrix}
    \equiv | b \rangle
\end{align}
where we defined the new $| b \rangle$ vector for this case allowing arbitrary force directions.  We can plug this $| b \rangle$ in Eq.~\eqref{EQ:kappa} to find the dipole stiffness for any given $(\theta,\phi)$, and represent the nontrivial term in $\kappa_b$ as
\begin{align}
    \eta_{b}
    &=
    \langle b | \cdot \displaystyle \sum_{i,j}^{N_{\textrm{SSS}}} |t_{\textrm{SSS},i}\rangle
    \lbrack (\mathbb{K}^{-1})_{ss} \rbrack^{-1}
    \langle t_{\textrm{SSS},j} | \cdot | b \rangle 
    \equiv \langle b | \mathbb{P} | b \rangle
    \label{EQ:eta_b}
    \\
    &=
    \begin{pmatrix}
     0&
     \cdots&
     \cos{\theta}&
     \sin{\theta}\cos{\phi}&
     \sin{\theta}\sin{\phi}&
     \cdots&
     0
    \end{pmatrix}
    \cdot \mathbb{P} \cdot \begin{pmatrix}
     0\\
     \vdots\\
     \cos{\theta}\\
     \sin{\theta}\cos{\phi}\\
     \sin{\theta}\sin{\phi}\\
     \vdots\\
     0\\
    \end{pmatrix}\\
    &=
    \begin{pmatrix}
     \cos{\theta}&
     \sin{\theta}\cos{\phi}&
     \sin{\theta}\sin{\phi}\\
    \end{pmatrix}
    \cdot \mathbb{P}_{b} \cdot
    \begin{pmatrix}
     \cos{\theta}\\
     \sin{\theta}\cos{\phi}\\
     \sin{\theta}\sin{\phi}\\
    \end{pmatrix}
\end{align}
where $\mathbb{P}$ represents the matrix in between $|b\rangle$ vectors in Eq.~\eqref{EQ:eta_b} and $\mathbb{P}_{b}$ is the  $3\times 3$ part of $\mathbb{P}$ associated with bond $b$.
%block diagonal matrix for $\mathbb{P}$.

Similarly,
\begin{align*}
    \langle b | \mathbb{K} \mathbb{C} \mathbb{Q}\mathbb{K}| b \rangle &=
    \begin{pmatrix}
     \cos{\theta}&
     \sin{\theta}\cos{\phi}&
     \sin{\theta}\sin{\phi}\\
    \end{pmatrix}
    \cdot {[\mathbb{K} \mathbb{C} \mathbb{Q}\mathbb{K}]}_{b} \cdot
    \begin{pmatrix}
     \cos{\theta}\\
     \sin{\theta}\cos{\phi}\\
     \sin{\theta}\sin{\phi}\\
    \end{pmatrix}\\
    \langle b | \mathbb{K} | b \rangle &=
    \begin{pmatrix}
     \cos{\theta}&
     \sin{\theta}\cos{\phi}&
     \sin{\theta}\sin{\phi}\\
    \end{pmatrix}
    \cdot {[\mathbb{K}]}_{b} \cdot
    \begin{pmatrix}
     \cos{\theta}\\
     \sin{\theta}\cos{\phi}\\
     \sin{\theta}\sin{\phi}\\
    \end{pmatrix}
\end{align*}
where ${[\mathbb{K} \mathbb{C} \mathbb{Q}\mathbb{K}]}_{b}$ and ${[\mathbb{K}]}_{b}$ are the $3\times 3$ parts of $\mathbb{K} \mathbb{C} \mathbb{Q}\mathbb{K}$ and $\mathbb{K}$ associated with bond $b$, respectively. 
%represents the $b$-th $3\times 3$ block diagonal matrix for the matrix calculated from $\mathbb{K} \mathbb{C} \mathbb{Q}\mathbb{K}$, and ${[\mathbb{K}]}_{b}$ s the $b$-th $3\times 3$ block diagonal matrix for $\mathbb{K}$.

As a result,
\begin{align*}
    \kappa_{b}
    &= \frac{ \langle b | \mathbb{K} \mathbb{C} \mathbb{Q}\mathbb{K} | b \rangle }{ \langle b|\mathbb{K}|b\rangle - \langle b | \displaystyle \sum_{i,j}^{N_{\textrm{SSS}}} |t_{\textrm{SSS},i}\rangle
    \lbrack (\mathbb{K}^{-1})_{ss} \rbrack^{-1}
    \langle t_{\textrm{SSS},j} | b \rangle}\\
    &=
    \frac{ \begin{pmatrix}
     \cos{\theta}&
     \sin{\theta}\cos{\phi}&
     \sin{\theta}\sin{\phi}\\
    \end{pmatrix}
    \cdot {[\mathbb{K} \mathbb{C} \mathbb{Q}\mathbb{K}]}_{b} \cdot
    \begin{pmatrix}
     \cos{\theta}\\
     \sin{\theta}\cos{\phi}\\
     \sin{\theta}\sin{\phi}\\
    \end{pmatrix} }{ 
    \begin{pmatrix}
     \cos{\theta}&
     \sin{\theta}\cos{\phi}&
     \sin{\theta}\sin{\phi}\\
    \end{pmatrix}
    \cdot \{ {[\mathbb{K}]}_{b}-\mathbb{P}_b \} \cdot
    \begin{pmatrix}
     \cos{\theta}\\
     \sin{\theta}\cos{\phi}\\
     \sin{\theta}\sin{\phi}\\
    \end{pmatrix}
    } .\\
\end{align*}

Minimizing $\kappa_{b}$ for a single bond $b$,
\begin{align}
    \kappa_{b,\textrm{min}}
    &= \min_{\theta, \phi}
    \left[
    \frac{ \begin{pmatrix}
     \cos{\theta}&
     \sin{\theta}\cos{\phi}&
     \sin{\theta}\sin{\phi}\\
    \end{pmatrix}
    \cdot {[\mathbb{K} \mathbb{C} \mathbb{Q}\mathbb{K}]}_{b} \cdot
    \begin{pmatrix}
     \cos{\theta}\\
     \sin{\theta}\cos{\phi}\\
     \sin{\theta}\sin{\phi}\\
    \end{pmatrix} }{ 
    \begin{pmatrix}
     \cos{\theta}&
     \sin{\theta}\cos{\phi}&
     \sin{\theta}\sin{\phi}\\
    \end{pmatrix}
    \cdot \{ {[\mathbb{K}]}_{b}-\mathbb{P}_b \} \cdot
    \begin{pmatrix}
     \cos{\theta}\\
     \sin{\theta}\cos{\phi}\\
     \sin{\theta}\sin{\phi}\\
    \end{pmatrix}
    }
    \right]
\end{align}
is an optimization problem with respect to the two variables $\theta$ and $\phi$. To solve for such optimization problems in our system, we used the Nelder-Mead simplex algorithm as described in Ref.~\cite{lagarias1998convergence}.  

By doing this minimization, we obtain  $(\theta,\phi)$ as the softest dipole stiffness direction for any bond $b$, which is typically transverse.  %The corresponding dipole stiffness is

\end{widetext}


%\bibliography{sss}

\end{document}